\documentclass[twocolumn]{aastex631}

\usepackage{float}
\usepackage{verbatim}
\usepackage[ ]{quoting} 
\usepackage{textcomp, gensymb}
\usepackage[mathlines]{lineno}
\usepackage{amsmath}
\usepackage{hyperref}
\usepackage{url}
\usepackage{witharrows}
\usepackage{xfrac}
\usepackage{multirow}

\usepackage{graphicx}
\usepackage[T1]{fontenc}
\usepackage{lmodern}
\usepackage{amssymb} 
\usepackage{comment}
\usepackage{enumerate}
\usepackage{bm}

\shorttitle{Carbon-deficient giants in the Kepler field}
\shortauthors{Maben et al.}

\begin{document}
\title{Asteroseismology sheds light on the origin of carbon-deficient red giants: likely merger products and linked to the Li-rich giants}

\author{Sunayana Maben}
\affil{CAS Key Laboratory of Optical Astronomy, National Astronomical Observatories, Chinese Academy of Sciences, Beijing 100101, Peoples Republic of China}
\affil{School of Astronomy and Space Science, University of Chinese Academy of Sciences, Beijing 100049, Peoples Republic of China}
\affil{Indian Institute of Astrophysics, 100ft road Koramangala, Bangalore, 560034, India}
  
\author{Simon W. Campbell}
\affil{School of Physics and Astronomy, Monash University, Clayton, Victoria, Australia}
\affil{ARC Centre of Excellence for Astrophysics in Three Dimensions (ASTRO-3D), Australia}

\author{Yerra Bharat Kumar}
\affil{Indian Institute of Astrophysics, 100ft road Koramangala, Bangalore, 560034, India}

\author{Bacham E. Reddy}
\affil{Indian Institute of Astrophysics, 100ft road Koramangala, Bangalore, 560034, India}

\author{Gang Zhao}
\affil{CAS Key Laboratory of Optical Astronomy, National Astronomical Observatories, Chinese Academy of Sciences, Beijing 100101, Peoples Republic of China}
\affil{School of Astronomy and Space Science, University of Chinese Academy of Sciences, Beijing 100049, Peoples Republic of China}

\correspondingauthor{Yerra Bharat Kumar and Gang Zhao}
\email{bharat.yerra@iiap.res.in; gzhao@nao.cas.cn}

\begin{abstract}
Carbon-deficient red giants (CDGs) are a peculiar class of stars that have eluded explanation for decades. We aim to better characterise CDGs by using asteroseismology (Kepler, TESS) combined with spectroscopy (APOGEE, LAMOST), and astrometry (Gaia). We discovered 15 new CDGs in the Kepler field, and confirm that CDGs are rare, being only $0.15\%$ of our background sample. Remarkably, we find that our CDGs are almost exclusively in the red clump (RC) phase. Asteroseismic masses reveal that our CDGs are primarily low-mass stars ($M \lesssim$ 2~M$_{\odot}$), in contrast to previous studies which suggested they are intermediate mass ($M = 2.5 - 5.0~\rm M_{\odot}$) based on HR diagrams. A very high fraction of our CDGs ($50\%$) are also Li-rich giants. We observe a bimodal distribution of luminosity in our CDGs, with one group having normal RC luminosity and the other being a factor of two more luminous than expected for their masses. We find demarcations in chemical patterns and luminosities which lead us to split them into three groups: (i) normal-luminosity CDGs, (ii) over-luminous CDGs, and (iii) over-luminous highly-polluted CDGs. We conclude that a merger of a helium white dwarf with an RGB star is the most likely scenario for the two groups of over-luminous stars. Binary mass-transfer from intermediate-mass AGB stars is a possibility for the highly-polluted over-luminous group. For the normal-luminosity CDGs, we cannot distinguish between core He-flash pollution or lower-mass merger scenarios. Due to the overlap with the CDGs, Li-rich giants may have similar formation channels.

\end{abstract}

\section{Introduction} \label{sec:introduction}

    When a star first ascends the red giant branch (RGB) the first dredge-up (FDU) occurs. In the FDU, the convective envelope moves inwards (in mass) and mixes material from the interior to the surface. This material has been exposed to H-burning on the main-sequence and therefore the surface composition of the star changes. There is good agreement between theory (e.g., \citealt{Iben1964,Iben1967,Dearborn1976,Dearborn1978}) and observation (e.g., \citealt{Lambert1977,Kjaergaard1982,Shetrone1993,Shetrone2019}) on this event. The main surface abundance changes are an increase in the $^4$He, $^{14}$N and $^{13}$C abundances, and a decrease in the $^{12}$C abundance by about 30\% \citep{Iben1984}.

    Contrary to this picture, 44 giants have been found to have extremely low carbon abundances (\citealt{Bidelman1951}; \citealt{Bidelman1973}; \citealt{Bond2019}), well below what is expected from FDU.  These stars are known as the weak G-band (wGb) stars, which are G and K giants whose spectra show very weak or absent G-band absorption of the CH molecule at 4300 \AA. Of the 44 wGb stars, only 29 stars have carbon abundances from high-resolution spectra (R $\approx$ 48,000 -  60,000; \citealt{Adamczak2013}; \citealt{Palacios2016}). When the high-resolution spectra of these peculiar stars were analyzed in detail, in addition to these stars being extremely carbon-deficient, their carbon isotopic ratios $\rm^{12}C/^{13}C$ were found to be close to the equilibrium value of 3 - 4. Also, N was found to be enhanced, and in some cases, they were found to be overabundant in Li and Na as well. These studies suggested that the wGb stars were probably intermediate-mass stars ranging in mass from about 2.5~$\rm M_{\odot}$ to 5.0~$\rm M_{\odot}$. Based on their position on the Hertzsprung–Russell diagram (HRD), many of them were shown to be in the sub-giant branch (SGB)/RGB phase and a few in the core He-burning phase (red clump; RC). The first dedicated spectroscopic survey for C-deficient stars was undertaken 50 years ago by \cite{Bidelman1973}. Recently, \cite{Bond2019} added 5 carbon-deficient giants (CDGs) based on spectroscopy to the initial list of CDGs. Since then no new CDGs based on spectroscopy have been identified.

    To this day, there is no consensus on why these giants have extremely low C and high N abundances. Some studies favor in-situ origin, that is internal nucleosynthesis with extra mixing in stars \citep{Adamczak2013}, and others favor an external origin such as pollution of their stellar atmospheres during the main sequence (MS) or pre-main sequence (PMS) by CN-processed material \citep{Palacios2016}.

    Two of the key factors that are required to decipher the possible origin of the carbon anomaly and its connection with other elements are the mass and evolutionary status of the wGb stars. Although there have been many attempts at determining these two stellar characteristics, the results so far have been inconclusive, since previous determinations have been based only on locations in the HRD (\citealt{Palacios2012,Adamczak2013,Palacios2016}; \citealt{Bond2019}). 
    
    In the current study, we aim to determine the mass and evolutionary status of these peculiar stars. For the masses we use asteroseismology combined with  astrometry, photometry, and spectroscopy. For the evolutionary phase, we will use the period spacing of the g-dominated mixed modes ($\Delta$P, $\Delta \Pi_{1}$) to determine if the wGb stars are in the RGB phase or He-core burning RC phase (\citealt{Bedding2011}; \citealt{Mosser2014}). These results, combined with chemical composition information, will hopefully allow us to better understand the origin of the carbon (and other) anomalies in these stars.

    To this end, in Section~\ref{sec:sample_selection} we build a catalog by conducting a large systematic search of wGb stars that have both asteroseismic data from the Kepler mission \citep{Borucki2010}, and spectroscopic data from the APOGEE survey \citep{Majewski2017}. 
    The results are presented in Section~\ref{sec:analysis_results}. We provide substantial discussion on the implications of our results in Section~\ref{sec:discussion}, and our conclusions are summarized in Section~\ref{sec:conclusion}.

\section{Sample selection} \label{sec:sample_selection}

\subsection{APOGEE-KEPLER cross-match}\label{sec:apogee_kepler_crossmatch}

    We take the spectroscopic data from data release 17 (DR17; \citealt{Abdurrouf2022}) of the Apache Point Observatory Galactic Evolution Experiment (APOGEE), which is a large-scale near-IR, high-resolution (R $\approx$ 22,500) survey. Stellar parameters and individual elemental abundances for up to 20 species were derived by the APOGEE Stellar Parameters and Chemical Abundances Pipeline (ASPCAP, \citealt{Garcia2016}) for 733,901 stars across the Milky Way.

    In the current study we focus on the Kepler field since we are interested in having asteroseismic constraints. We explore the entire APOGEE sample in a separate study (Maben et al. 2023, in preparation). We use the Kepler stellar properties catalog  by \cite{Mathur2017}. This catalog is based on Kepler data release 25, which contains data from Quarters~1 to 17. It is the largest catalog of stars observed by the Kepler mission with 197,096 individual targets.

    To begin with, we cross-matched these two catalogs, which yielded a sample of 23,129 stars (APOGEE-KEPLER sample). Secondly, we checked how many of these stars have global asteroseismic parameters derived from precision Kepler light curves in the literature (\citealt{Stello2013}; \citealt{Mosser2014}; \citealt{Vrard2016}; \citealt{Serenelli2017}; \citealt{Yu2018}; \citealt{Pinsonneault2018}; \citealt{Li2020}). We found 11,099 stars (i.e., 48\% of the APOGEE-KEPLER sample) that matched this criterion. 

    Further, we constrain our sample to have good-quality APOGEE spectroscopic data. The \texttt{ASPCAPFLAG} is used to flag potential issues with an observation and/or with specific stellar parameters. Hence, we remove stars with flags \texttt{STAR\_BAD} or \texttt{STAR\_WARN}. Since we also require reliable metallicity and carbon abundances, we limit our sample to stars that have \texttt{FE\_H\_FLAG=0} and \texttt{C\_FE\_FLAG=0} \citep{Jonsson2020}.

    We restrict our sample to disk metallicities ([Fe/H]~$\geq -0.8$~dex; \citealt{Tomkin1995}). This is consistent with the known wGb stars. Carbon is more complex at low metallicities since there is a significant dispersion in abundances  (e.g., \citealt{Romano2019}). In addition, there is a known carbon-depletion mechanism -- deep-mixing in red giants -- that increases the carbon depletion rate at low metallicities (e.g., \citealt{Martell2008}; \citealt{Gerber2019}). These constraints give us 10,674 stars (i.e., 46\% of the APOGEE-KEPLER sample). 

    One of the primary purposes of this study is to determine the evolutionary phase of CDGs. We therefore, select stars that have their evolutionary status positively determined in the different asteroseismic studies mentioned above. Applying this criterion we obtain a final common sample of 10,180 stars (i.e., 44\% of the APOGEE-KEPLER sample).

    \begin{figure*}[htp]
    \centering
    \includegraphics[width=1.\linewidth]{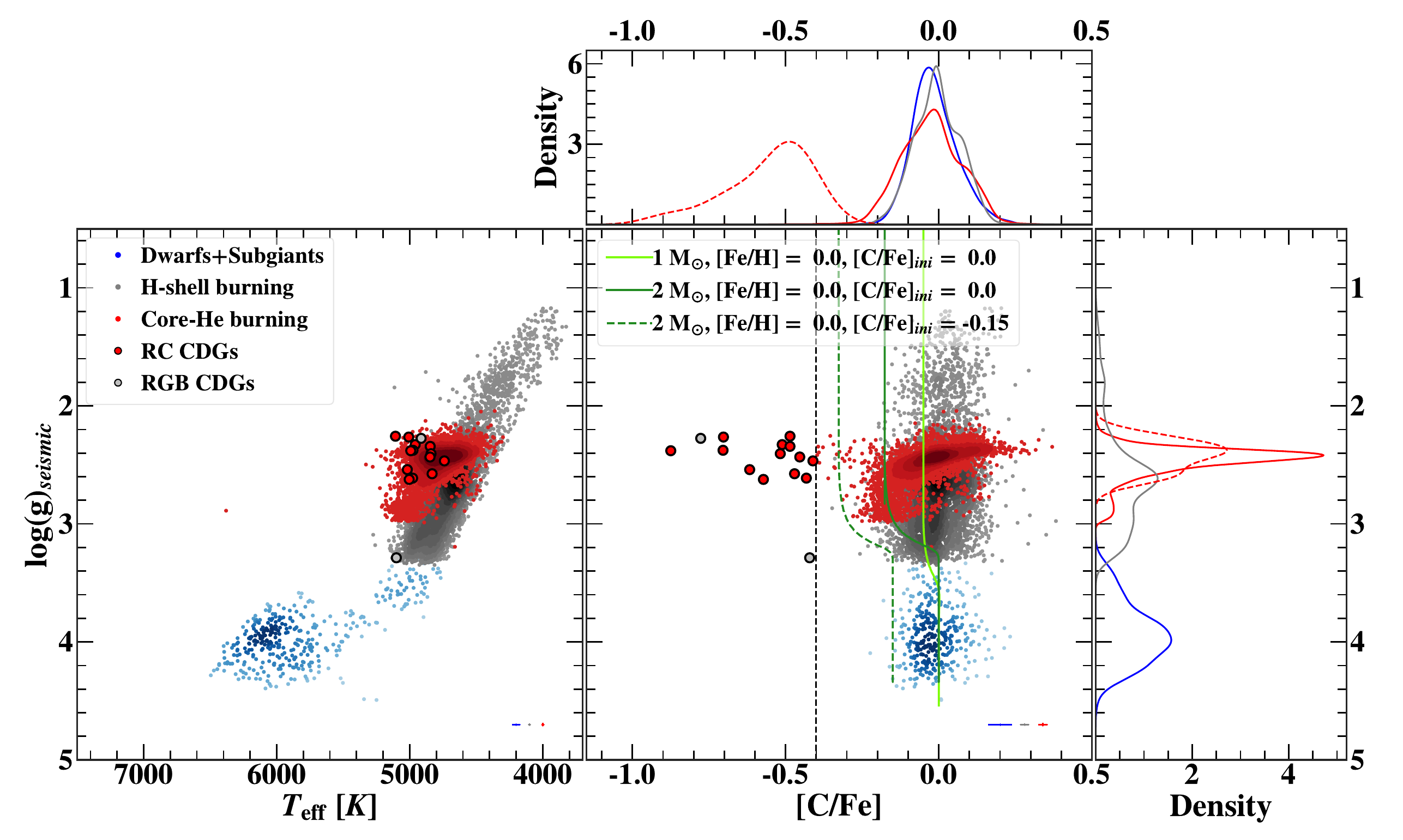} 
    \caption{Hertzsprung–Russell diagram (left) and log(g)$_{\rm seis}$ as a function of [C/Fe] (right). Our APOGEE-KEPLER common sample forms the background in a color scale that represents the number density of stars; darker colors indicating higher density: Dwarfs and sub-giants (blue dots), inert He-core giants ascending the RGB for the first time (grey dots) and He-core burning giants (red dots). Kernel density histograms are included to show the log(g) and [C/Fe] distributions for various parts of the sample, with the color scheme following the background sample, and the dashed histogram showing our final C-deficient sample. We superimpose model tracks in the log(g)-[C/Fe] plane for comparison (see text for model details). Starting with an initial [C/Fe]~$=0.0$, the models show a reduction of [C/Fe] by $\approx 0.05$ to $0.2$~dex, due to FDU dilution. As some dwarf stars have lower [C/Fe], which may reflect a lower initial abundance, we offset the 2~M$_{\odot}$ model (which has the larger C-depletion) by $-0.15$ dex to match the lower envelope of those stars. This indicates the expected lower limit of [C/Fe] for later phases of evolution (dashed model line). The vertical dashed line at [C/Fe]~$= -0.4$~dex denotes the cut we use to define carbon-deficient stars (see text for details). The larger symbols (filled circles) show our carbon-deficient sample. Large red circles are RC stars and the two large grey circles are RGB stars. Formal error bars are shown on the bottom right end of each subplot. The top and right panel of the right subplot shows kernel density histograms for the normal giants (solid lines) and RC CDGs (dashed red line). We use the solar abundance of C as derived by \cite{Grevesse2007}.}
    \label{fig:fig1}
    \end{figure*}

    \subsection{Defining and identifying carbon-deficient stars} \label{sec:definition}

        Figure~\ref{fig:fig1} shows the APOGEE-KEPLER common sample that made it through the quality criteria discussed in Section \ref{sec:apogee_kepler_crossmatch}. In the left panel of the figure, we see that we have a sample that covers all the evolutionary phases, from dwarfs and sub-giants to H-shell burning stars and core He-burning (CHeB) stars. The distribution of the carbon abundance for the sample is shown in the central panel. It can immediately be seen that there are very few outliers, with low [C/Fe]. 

        Here we want to make a quantitative definition of carbon-deficient stars. To guide us, we use stellar models, which illustrate the theoretical expectations for the carbon surface abundances. We calculated a series of models using the MESA stellar code (\citealt{Paxton2011,Paxton2019}; version 12778). RGB mass loss was modeled using the \cite{Reimers1975} formula ($\eta = 0.3$). We used the standard MESA nuclear network (`basic.net'), standard equation of state (see \citealt{Paxton2019} for details), and $\alpha_{MLT} = 2.0$. Convective boundary locations were based on the Schwarzschild criterion, extended with exponential overshoot (\citealt{Herwig1997}) during core helium burning ($f_{OS} = 0.001$, following \citealt{Constantino2017}). Models were run from the pre-main sequence to the start of the thermally pulsing-asymptotic giant branch (TP-AGB) phase, so as to cover the CHeB phase for which we have many stars in our sample. In Figure~\ref{fig:fig1} we show two of our models, having masses of 1~M${_{\odot}}$ and 2~M${_{\odot}}$. These model masses are representative of the bulk of the background observational sample. There is a metallicity distribution in our sample with [Fe/H] ranging from $-0.8$ to $+0.5$~dex and peaking at [Fe/H]~$= 0.0$~dex. Our models showed that metallicity has a much smaller effect than mass on [C/Fe], so we only show [Fe/H]~$= 0.0$ models for clarity. As can be seen in Figure~\ref{fig:fig1}, the models start at [C/Fe]~$= 0.0$ then deplete C around log(g)~$=3.2-3.5$, as the convective envelope deepens during FDU early on the RGB. By log(g)~$\approx 2.8-3.2$, the surface C abundances have reached their minima, with a total [C/Fe] reduction of $0.05$~dex and $0.18$~dex for the 1~M${_{\odot}}$ and 2~M${_{\odot}}$ models, respectively. From then on the C remains constant.

        The initial C abundance of a star will affect its future surface abundance. Our sample includes dwarf stars that have abundances as low as [C/Fe]~$\simeq  -0.15$~dex. To roughly account for stars with lower initial C we offset the 2~M$_{\odot}$ model track (which has the largest C depletion) by $-0.15$ dex (similar to \citealt{Mishenina2006} and \citealt{Tautvaisiene2010}). Here, the carbon depletion reaches as low as [C/Fe]~$\simeq -0.33$~dex.  Based on this, and the observed distribution of the low-carbon tail, we make a conservative cut at [C/Fe]~$ = -0.4$~dex, below which we consider all stars as carbon-deficient (vertical dashed line in Figure~\ref{fig:fig1}). With this definition, we find a total of 15 carbon-deficient stars, $0.15\%$ of our final APOGEE-KEPLER sample. This highlights how rare carbon-deficient stars are. We will be referring to these stars as the `CDGs'. All stars in our CDG sample have known evolutionary phases. In Figure~\ref{fig:fig1}, we indicate their respective evolutionary phases using different symbols. Thirteen of our CDGs are RC stars, and two are RGB stars (although we later re-classify one of those as RC; see Section~\ref{sec:luminosity}).

\section{Analysis and Results} \label{sec:analysis_results}
\subsection{Evolutionary phase: Seismic diagram} \label{sec:seismic_diagram}

\begin{figure}[htp]
    \centering
    \includegraphics[width=0.9\linewidth]{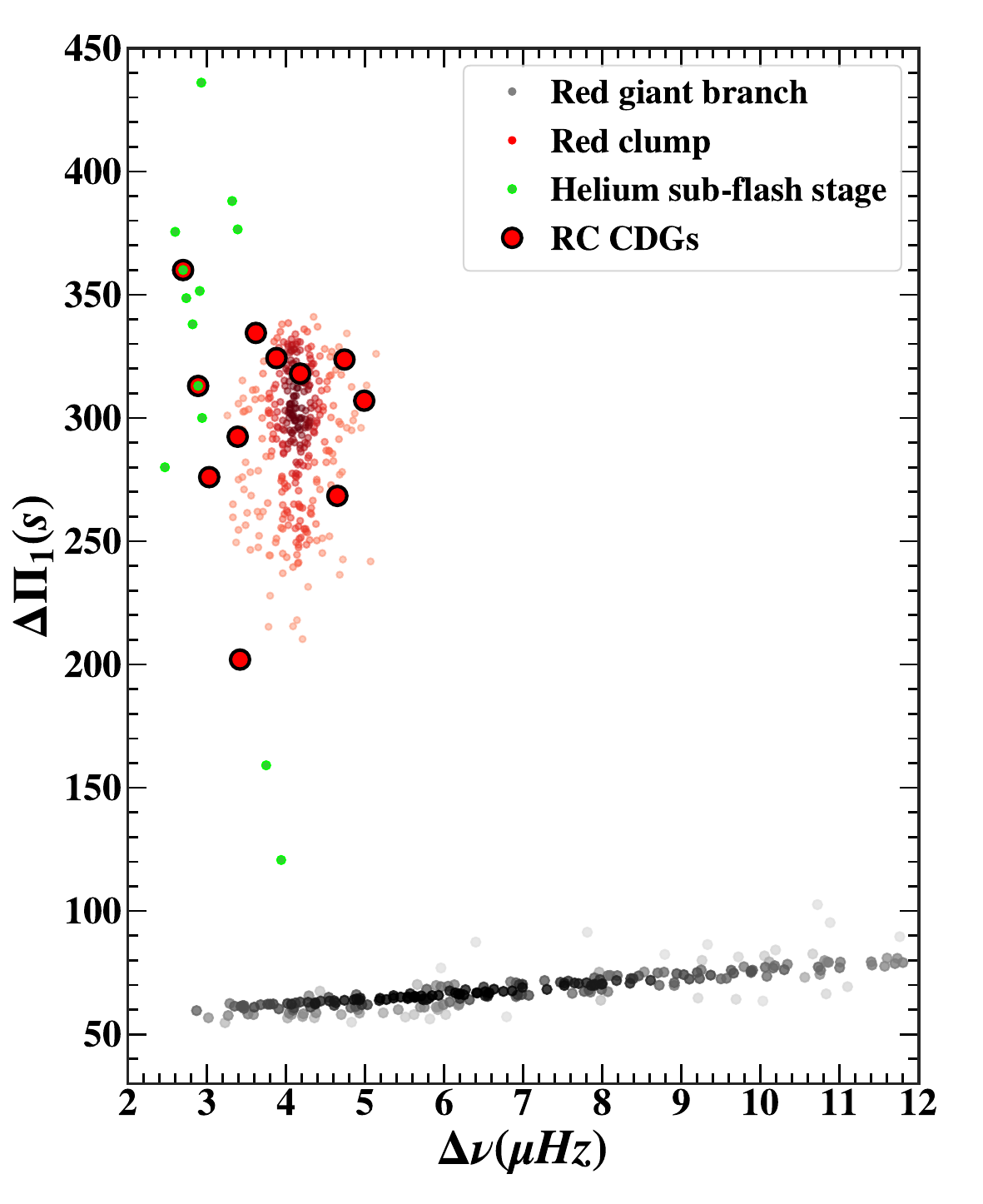} 
    \caption{Seismic diagram of our RC CDGs which have known $\Delta\Pi_{1}$ and $\Delta \nu$ (11 stars; large filled red circles). Giants classified based on asteroseismic analysis form the background from  \cite{Mosser2014} (small filled circles; see key), in a color scale that represents the number density of stars; darker colors indicate higher densities.}
    \label{fig:fig2}
    \end{figure}

    It has been shown that RC and RGB stars can be separated by considering measurements of their mixed-mode period spacing $\Delta$P \citep{Bedding2011} or equivalently, the asymptotic g-mode period spacing $\Delta\Pi_{1}$ \citep{Mosser2014}. Red giants with $\Delta\Pi_{1}\geq 80$ seconds are RC stars and the giants in the very narrow strip (Figure~\ref{fig:fig2}) with $\Delta\Pi_{1} \leq 80$~seconds are RGB stars \citep{Vrard2016}. 

    Out of our 15 CDGs, 6 stars have both the asteroseismic parameters $\Delta\Pi_{1}$ and $\Delta \nu$ from \cite{Mosser2014}, and 5 stars have these parameters from \cite{Vrard2016}\footnote{The single RGB star (see Section~\ref{sec:luminosity}) does not have a reported $\Delta\Pi_{1}$, but it is clearly of too low luminosity to be a RC star.}. 
    In Figure~\ref{fig:fig2} we plot these 11 RC stars (large symbols) over the background sample of \cite{Mosser2014}. We have limited the background sample to match our RC sample mass range of 1.0~M$_{\odot}$ $\leq M \leq$ 2.5~M$_{\odot}$ (Section~\ref{sec:masses}) for better comparison.

    As seen in Figure~\ref{fig:fig2}, \textit{all} of our CDGs with $\Delta\Pi_{1}$ have large values of this parameter. They also have small values of $\Delta \nu$ ($<$ 5~$\mu$Hz). They clearly occupy the He-core burning phase region of the $\Delta\Pi_{1}$-$\Delta \nu$ diagram. Two of the CDGs are classified as Helium subflashing stars by \cite{Mosser2014} (see discussion in Section~\ref{sec:olrc_subflash}).

\subsection{Luminosities} \label{sec:luminosity}

    Luminosities of the CDGs were determined using the standard formula\footnote{log($L/L_{\odot}$) = $-$0.4 [$V_{0}$ $-$ $(m - M)_{0}$ $+$ $BC$ $-$ $M_{bol,\odot}$]}. Distances were taken from the catalog of \cite{BailerJones2021}. The visual magnitudes and their errors were estimated from the color-color transformations given in \cite{Riello2021} that relate the Gaia DR3 photometric system to the Johnson-Cousins system \citep{Stetson2000}. We used the \cite{Green2019} three-dimensional dust map to estimate the reddening for each of the stars. We applied bolometric corrections following \cite{Alonso1999}, adopting  errors in $T_{\rm eff}$ and [Fe/H] of $50$~K and $0.05$~dex, respectively. Our derived luminosities are listed   in Table \ref{tab:table1}.

    All of our stars, except one (KIC~8352953), have luminosities consistent with being RC stars (Table~\ref{tab:table1}, also see Kiel diagram in Figure~\ref{fig:fig1}). KIC~8352953 has a luminosity that coincides with the lower-RGB phase and is classified as an RGB star in \cite{Yu2018}. All except one of the stars (KIC~8222189) that have RC luminosities are classified as RC stars by asteroseismic studies (\citealt{Stello2013}; \citealt{Mosser2014}; \citealt{Vrard2016}; \citealt{Yu2018}; \citealt{Pinsonneault2018}). This one star has a stellar evolutionary phase of RGB (or AGB) in \cite{Pinsonneault2018}, is unclassified in \cite{Yu2018} and \cite{Elsworth2017}, but is classified as a RC star by \cite{Ting2018}. Given that KIC~8222189 has a luminosity and surface gravity that is consistent with the core He-burning phase, we classify it as a RC star from now on, but note that a $\Delta P$ measurement would be required to be certain. In summary, our CDG sample has one RGB star and 14 RC stars.

    \begin{figure}[htp]
    \centering
    \includegraphics[scale=0.5]{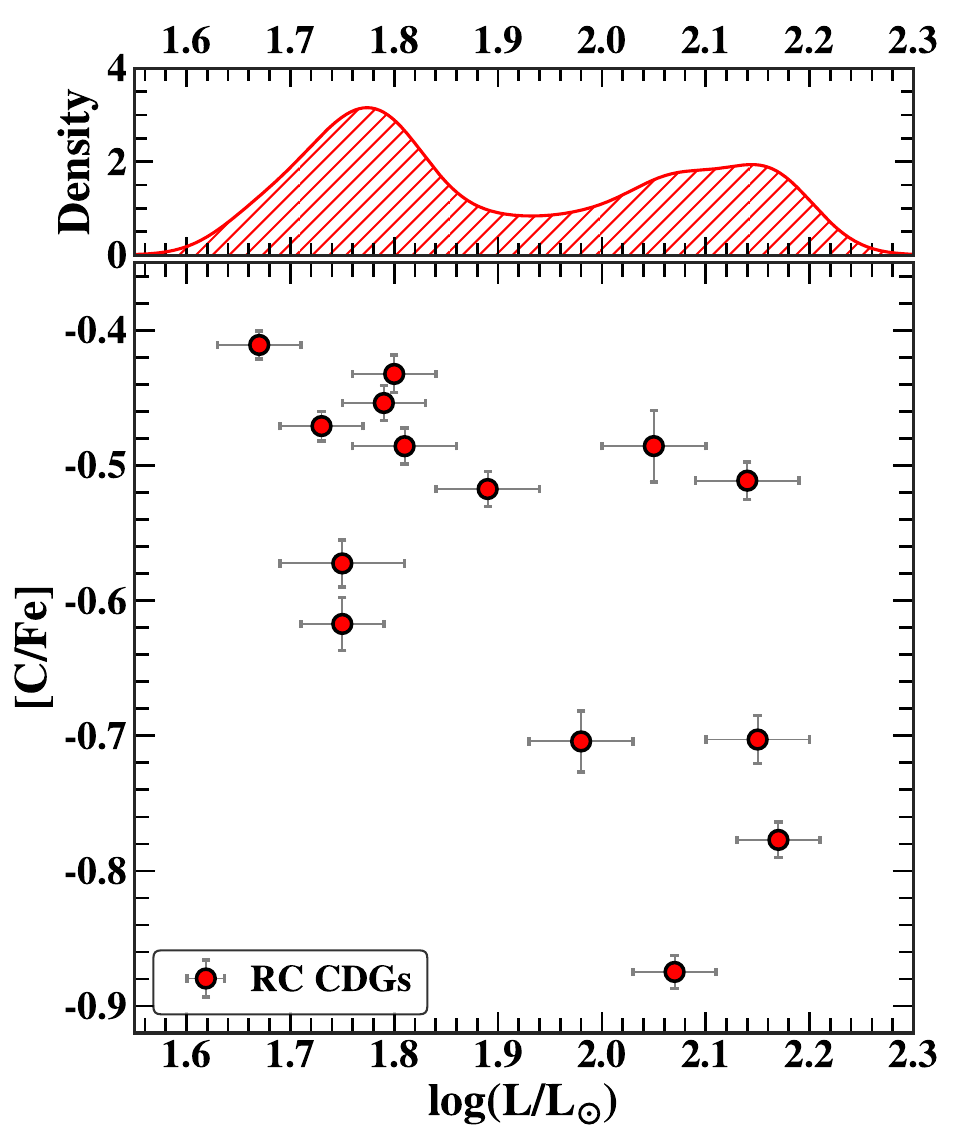}
    \caption{Carbon abundance as a function of $\log(L/L_{\odot})$ for our RC CDGs. As can be seen in the Gaussian kernel density histogram above, the luminosity distribution is bimodal. This bimodality is discussed in detail in Section~\ref{sec:bimodality}. We use the solar abundance of C as derived by \cite{Grevesse2007}.}
    \label{fig:fig3}
    \end{figure}

    Looking at the distribution of the luminosities, we find there is a bimodality (see Figure~\ref{fig:fig3}), with one group having higher luminosity ($\log(L/L_{\odot}) = 2.1 \pm 0.1$~dex) than the other ($\log(L/L_{\odot}) = 1.8\pm 0.1$~dex). Interestingly the brighter RC stars tend to have the lowest C abundances (see Figures~\ref{fig:fig3} and \ref{fig:fig10}). The bi-modality is also present in log(g) but is not as clear (Figure~\ref{fig:fig1}). We discuss this bimodality in detail in Section~\ref{sec:bimodality}.

    \begin{table*}[]
    \begin{rotatetable*}
    \begin{center}

    \caption{Atmospheric, asteroseismic, kinematic parameters along with the CNO abundance ratios and mass estimates from equations (\ref{eq:mass1}) - (\ref{eq:mass5}) of the CDGs. We have adopted the errors in $T_{\rm eff}$ and [Fe/H] as 50 K and 0.05 dex respectively for all the stars. $M_{avg}$ is the average of the asteroseismic mass equations (\ref{eq:mass1}), (\ref{eq:mass3}), and (\ref{eq:mass4}).}
    \label{tab:table1}
    \hspace*{-26em}
    \resizebox{1.63\linewidth}{!}{%
    \begin{tabular}{lrrrrrrrlcrrrrrlrrrl}
    \hline
    \multicolumn{1}{c}{KIC} & \multicolumn{1}{c}{$T_{\rm eff}$} & \multicolumn{1}{c}{{[}Fe/H{]}} & \multicolumn{1}{c}{log(g)$_{\rm seis}$$^{c}$} & \multicolumn{1}{c}{$\log(L/L_{\odot})$} & \multicolumn{1}{c}{$\nu_{\rm max}$$^{c}$} & \multicolumn{1}{c}{$\Delta \nu$} & \multicolumn{1}{c}{$\Delta\Pi_{1}$} & \multicolumn{1}{c}{{[}C/Fe{]}} & \multicolumn{1}{c}{{[}N/Fe{]}} & \multicolumn{1}{c}{{[}O/Fe{]}} & \multicolumn{1}{c}{M$_{1}$} & \multicolumn{1}{c}{M$_{2}$} & \multicolumn{1}{c}{M$_{3}$} & \multicolumn{1}{c}{M$_{4}$} & \multicolumn{1}{c}{M$_{avg}$} & \multicolumn{1}{c}{M$_{5}$}  & \multicolumn{1}{c}{R$_{\rm seis}$} & \multicolumn{1}{c}{b} & \multicolumn{1}{c}{Z}     \\
    \multicolumn{1}{c}{}    & \multicolumn{1}{c}{(K)}       & \multicolumn{1}{c}{}           & \multicolumn{1}{c}{}            & \multicolumn{1}{c}{}                   & \multicolumn{1}{c}{($\mu$Hz)}      & \multicolumn{1}{c}{($\mu$Hz)}   & \multicolumn{1}{c}{(s)}       & \multicolumn{1}{c}{}           & \multicolumn{1}{c}{}           & \multicolumn{1}{c}{}           & \multicolumn{1}{c}{(M$_{\odot}$)}         & \multicolumn{1}{c}{(M$_{\odot}$)}   & \multicolumn{1}{c}{(M$_{\odot}$)}      & \multicolumn{1}{c}{(M$_{\odot}$)}         & \multicolumn{1}{c}{(M$_{\odot}$)}         & \multicolumn{1}{c}{(M$_{\odot}$)}     & \multicolumn{1}{c}{(R$_{\odot}$)}     & \multicolumn{1}{c}{(deg)}  & \multicolumn{1}{c}{(kpc)}  \\
    \hline
    \hline
    5881715                 & 4840                       & -0.12                          & 2.41$\pm$0.02                              & 1.89$\pm$0.05                          & 30.90                               & 3.42$^{a}$                             & 202$^{a}$                           & -0.52                          & 0.53                           & -0.02                          & 1.91$\pm$0.22                             & 1.26$\pm$1.86                             & 1.44$\pm$0.18                             & 1.76$\pm$0.17     & 1.70                        & 1.35$\pm$0.23  &  14.46$\pm$0.63                      & 10.15                 & 0.33                                           \\
    8879518                 & 4832                       & 0.08                           & 2.58$\pm$0.01                              & 1.73$\pm$0.04                          & 46.18                               & 4.65$^{b}$                             & 268$^{b}$                         & -0.47                          & 0.61                           & -0.01                          & 1.87$\pm$0.14                             & 1.35$\pm$2.12                             & 1.50$\pm$0.15                             & 1.74$\pm$0.11  & 1.70                           & 2.11$\pm$0.32                        &  11.66$\pm$0.35    & 14.46                 & 0.32                                    \\
    4071012                 & 4992                       & 0.07                           & 2.38$\pm$0.01                              & 2.07$\pm$0.04                          & 29.20                               & 3.03$^{b}$                             & 276$^{b}$                           & -0.87                          & 0.68                           & -0.11                          & 2.74$\pm$0.25                             & 1.53$\pm$0.84                             & 1.85$\pm$0.19                             & 2.43$\pm$0.18   & 2.34                          & 2.81$\pm$0.43 &  17.65$\pm$0.70                          & 8.18                  & 0.27                                         \\
    3355015                 & 4846                       & -0.15                          & 2.34$\pm$0.02                              & 1.81$\pm$0.05                          & 26.98                               & 3.39$^{a}$                             & 292$^{a}$                         & -0.49                          & 0.57                           & 0.03                           & 1.31$\pm$0.14                             & 0.94$\pm$1.31                             & 1.04$\pm$0.13                             & 1.22$\pm$0.11    & 1.19                         & 1.06$\pm$0.18  & 12.76$\pm$0.51                        & 8.15                  & 0.29                                      \\
    3736289                 & 4978                       & -0.10                          & 2.61$\pm$0.01                              & 1.80$\pm$0.04                          & 49.94                               & 4.99$^{a}$                             & 307$^{a}$                           & -0.43                          & 0.52                           & -0.04                          & 1.87$\pm$0.17                             & 1.65$\pm$0.98                             & 1.72$\pm$0.18                             & 1.82$\pm$0.14      & 1.80                       & 1.60$\pm$0.24   & 11.16$\pm$0.37                          & 12.78                 & 0.28                                      \\
    5446927                 & 5107                       & -0.74                          & 2.26$\pm$0.01                              & 2.05$\pm$0.05                          & 21.88                               & 2.89$^{a}$                             & 313$^{a}$                           & -0.49                          & 0.47                           & 0.15                           & 1.46$\pm$0.12                             & 1.12$\pm$0.39                             & 1.22$\pm$0.15                             & 1.38$\pm$0.10    & 1.36                         & 1.34$\pm$0.22  & 14.80$\pm$0.51                          & 11.36                 & 0.50                                        \\
    4667911                 & 4740                       & 0.08                           & 2.47$\pm$0.01                              & 1.67$\pm$0.04                          & 36.37                               & 4.18$^{b}$                             & 318$^{b}$                           & -0.41                          & 0.53                           & -0.03                          & 1.34$\pm$0.13                             & 1.00$\pm$3.43                             & 1.10$\pm$0.11                             & 1.26$\pm$0.10    & 1.23                         & 0.98$\pm$0.15  & 11.19$\pm$0.42                           & 8.34                  & 0.15                                         \\
    5000307                 & 5018                       & -0.29                          & 2.54$\pm$0.01                              & 1.75$\pm$0.04                          & 42.16                               & 4.74$^{a}$                             & 324$^{a}$                         & -0.62                          & 0.63                           & -0.01                          & 1.39$\pm$0.09                             & 1.20$\pm$0.57                             & 1.26$\pm$0.13                             & 1.35$\pm$0.07     & 1.33                        & 1.49$\pm$0.23   & 10.43$\pm$0.26                         & 13.20                 & 0.33                                         \\
    11971123                & 4848                       & -0.13                          & 2.43$\pm$0.01                              & 1.79$\pm$0.04                          & 32.47                               & 3.88$^{b}$                             & 324$^{b}$                         & -0.45                          & 0.52                           & 0.05                           & 1.33$\pm$0.13                             & 1.14$\pm$1.58                             & 1.20$\pm$0.12                             & 1.29$\pm$0.10        & 1.27                     & 1.37$\pm$0.21  & 11.74$\pm$0.45                           & 13.04                 & 0.31                                         \\
    8110538                 & 4975                       & -0.59                          & 2.38$\pm$0.02                              & 1.98$\pm$0.05                          & 28.77                               & 3.62$^{b}$                             & 334$^{b}$                         & -0.70                          & 0.80                           & 0.49                           & 1.29$\pm$0.14                             & 1.62$\pm$1.00                             & 1.50$\pm$0.19                             & 1.35$\pm$0.12       & 1.38                      & 1.12$\pm$0.19  & 12.19$\pm$0.52                           & 9.74                  & 0.62                                         \\
    2423824                 & 5007                       & -0.39                          & 2.26$\pm$0.01                              & 2.15$\pm$0.05                          & 22.02                               & 2.70$^{a}$                             & 360$^{a}$                           & -0.70                          & 0.63                           & 0.08                           & 1.88$\pm$0.18                             & 1.56$\pm$0.81                             & 1.66$\pm$0.20                             & 1.81$\pm$0.14     & 1.79                        & 1.71$\pm$0.29  &  16.85$\pm$0.67                           & 13.59                 & 0.49                                         \\
    8352953                 & 5101                       & -0.18                          & 3.29$\pm$0.01                              & 1.10$\pm$0.04                          & 230.27                              & 16.74$^{c}$                            &                               & -0.42                          & 0.63                           & 0.24                           & 1.46$\pm$0.07                             & 1.45$\pm$0.48                             & 1.45$\pm$0.15                             & 1.46$\pm$0.07          & 1.45                   & 1.66$\pm$0.25   &  4.56$\pm$0.08                         & 16.27                 & 0.26                                        \\
    7848354                 & 5004                       & -0.08                          & 2.62$\pm$0.02                              & 1.75$\pm$0.06                          & 51.74                               & 4.67$^{c}$                             &                               & -0.57                          & 0.59                           & 0.03                           & 2.66$\pm$0.26                             & 1.19$\pm$0.64                             & 1.56$\pm$0.23                             & 2.26$\pm$0.20         & 2.16                    & 1.67$\pm$0.31  & 13.05$\pm$0.44                          & 7.59                  & 0.42                                         \\
    4830861                 & 4959                       & -0.12                          & 2.33$\pm$0.01                              & 2.14$\pm$0.05                          & 25.94                               & 2.96$^{c}$                             &                               & -0.51                          & 0.56                           & 0.01                           & 2.06$\pm$0.18                             & 1.94$\pm$1.29                             & 1.98$\pm$0.24                             & 2.03$\pm$0.15      & 2.02                       & 2.67$\pm$0.45    &  16.26$\pm$0.59                       & 11.81                 & 0.43                             \\     
    8222189                 & 4914                       & -0.09                          & 2.28$\pm$0.02                              & 2.17$\pm$0.04                          & 23.02                               & 2.69$^{c}$                             &                               & -0.78                          & 0.65                           & -0.06                           & 2.07$\pm$0.38                             & 1.88$\pm$1.62                             & 1.94$\pm$0.22                             & 2.03$\pm$0.30  & 2.02                           & 2.43$\pm$0.37                &  17.37$\pm$1.16           & 15.57                  & 0.35                                         \\
    \hline
    \end{tabular}%
    
    }
    \begin{quoting}
          {\small {\bf Note:} $^{a}$   \cite{Mosser2014};
          $^{b}$  \cite{Vrard2016};
          $^{c}$   \cite{Yu2018}}
    \end{quoting}
    \end{center}
\end{rotatetable*}
\end{table*}

\subsubsection{Masses} \label{sec:masses}

    The masses of the CDGs were determined using several combinations of the seismic ($\nu_{max}$, $\Delta \nu$) and non-seismic parameters ($T_{\rm eff}$, $L/L_{\odot}$, $g$) that were available. The global seismic parameters are correlated with fundamental stellar properties; $\nu_{max} \propto g T_{\rm eff}^{-1/2}$ and $\Delta \nu \propto \rho^{-1/2}$, where $g$ is the surface gravity and $\rho$ is the mean density of the star (\citealt{Ulrich1986}; \citealt{Brown1991}; \citealt{Kjeldsen1995}). Combining these relations with the Stefan–Boltzmann luminosity law, L $\propto$ R$^{2}$ $T_{\rm eff}^{4}$, allows the derivation of the four seismic mass equations (Equations~(\ref{eq:mass1})-(\ref{eq:mass4}); e.g., \citealt{Miglio2016,Howell2022}). Equation (\ref{eq:mass5}) is a standard, but non-seismic mass formula, which uses the photometric and spectroscopic parameters of a star. We determined the masses of the CDGs using all of these 5 mass equations:

    \begin{linenomath}
    \begin{align}
    \left( \frac{M}{M_\odot}\right)  & \simeq \left( \frac{\nu_{max}}{\nu_{max,\odot}}\right)^{3} \left( \frac{\Delta \nu}{\Delta \nu_{\odot}}\right)^{-4} \left( \frac{T_{\rm eff}}{T_{\rm eff,\odot}}\right)^{3/2} \label{eq:mass1}\\
    \left( \frac{M}{M_\odot}\right)  & \simeq \left( \frac{\Delta \nu}{\Delta \nu_{\odot}}\right)^{2} \left( \frac{L}{L_{\odot}}\right)^{3/2} \left( \frac{T_{\rm eff}}{T_{\rm eff,\odot}}\right)^{-6} \label{eq:mass2}\\
    \left( \frac{M}{M_\odot}\right) & \simeq \left(\frac{\nu_{max}}{\nu_{max,\odot}}\right) \left(\frac{L}{L_{\odot}}\right) \left(\frac{T_{\rm eff}}{T_{\rm eff,\odot}}\right)^{-7/2} \label{eq:mass3}\\
    \left( \frac{M}{M_\odot}\right) & \simeq \left(\frac{\nu_{\rm max}}{\nu_{max,\odot}}\right)^{12/5} \left(\frac{\Delta \nu}{\Delta \nu_{\odot}}\right)^{-14/5} \left(\frac{L}{L_{\odot}}\right)^{3/10} \label{eq:mass4}\\
    \left( \frac{M}{M_\odot}\right) & \simeq \left(\frac{L}{L_{\odot}}\right) \left(\frac{g}{g_{\odot}}\right) \left(\frac{T_{\rm eff}}{T_{\rm eff,\odot}}\right)^{-4}\label{eq:mass5}
    \end{align} 
    \end{linenomath}

    Here, we use $\nu_{max,\odot} = 3090 \pm 30~\mu$Hz, $\Delta \nu_{\odot} = 135.1 \pm 0.1~\mu$Hz (both from \citealt{Huber2011}), $T_{\rm eff,\odot} = 5777$~K, and $\log(g)_{\odot}=4.44$ (both from \citealt{Morel2014}) as our adopted solar values. 

    For Equations (\ref{eq:mass1}), (\ref{eq:mass2}) and (\ref{eq:mass4}), it is known that one needs to apply a correction to the observed $\Delta \nu$ for use in the scaling relations (e.g., \citealt{White2011, Miglio2012}). We calculate $\Delta \nu$ corrections using  \texttt{ASFGRID} \citep{Sharma2016} and apply these to the $\Delta \nu$ values of the CDGs obtained from \cite{Mosser2014} and \cite{Vrard2016} only, as \cite{Yu2018} provides corrected $\Delta \nu$.

    \begin{figure}[htp]
    \centering
    \includegraphics[width=\linewidth]{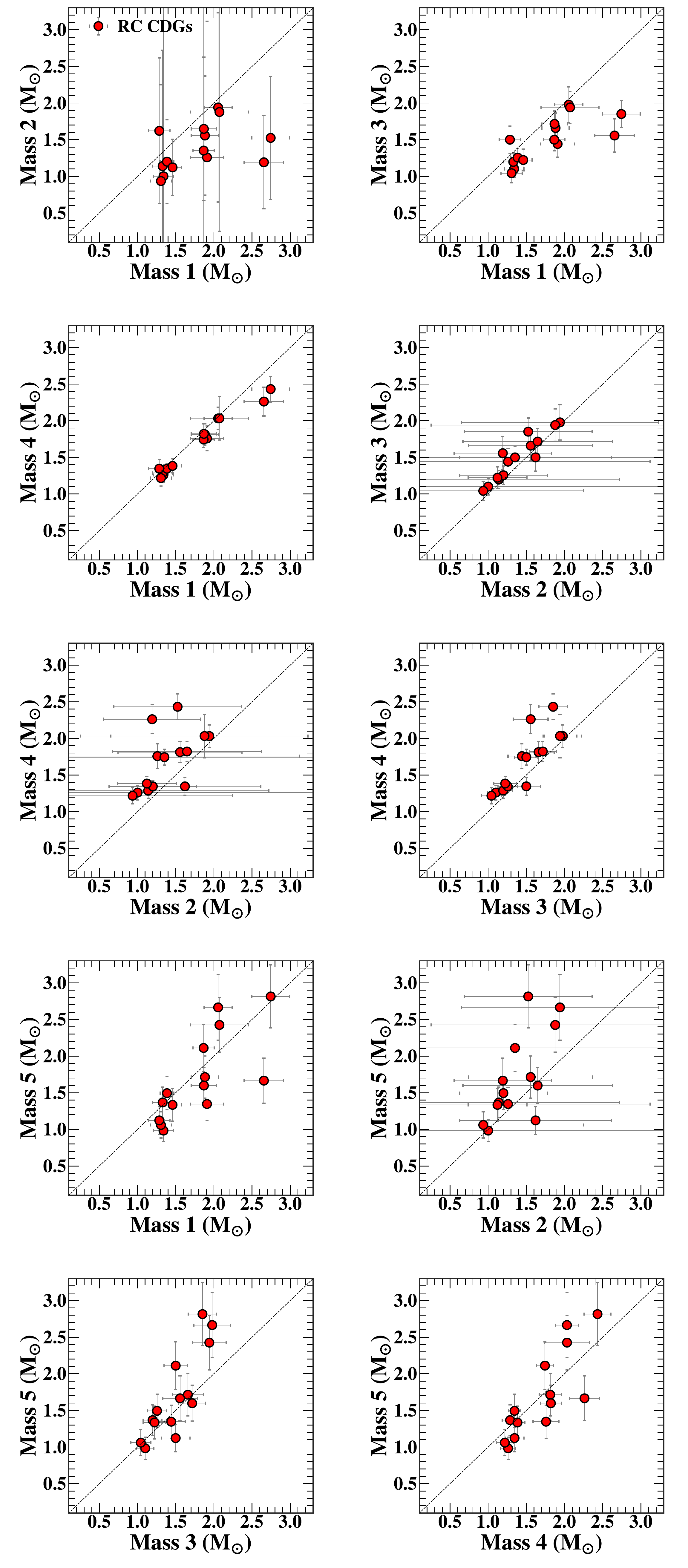}
    \caption{Comparison of various mass  estimates of the RC CDGs using Equations (\ref{eq:mass1})-(\ref{eq:mass5}).}
    \label{fig:fig4}
    \end{figure}

    \begin{figure}[htp]
    \centering
    \includegraphics[width=0.65\columnwidth]{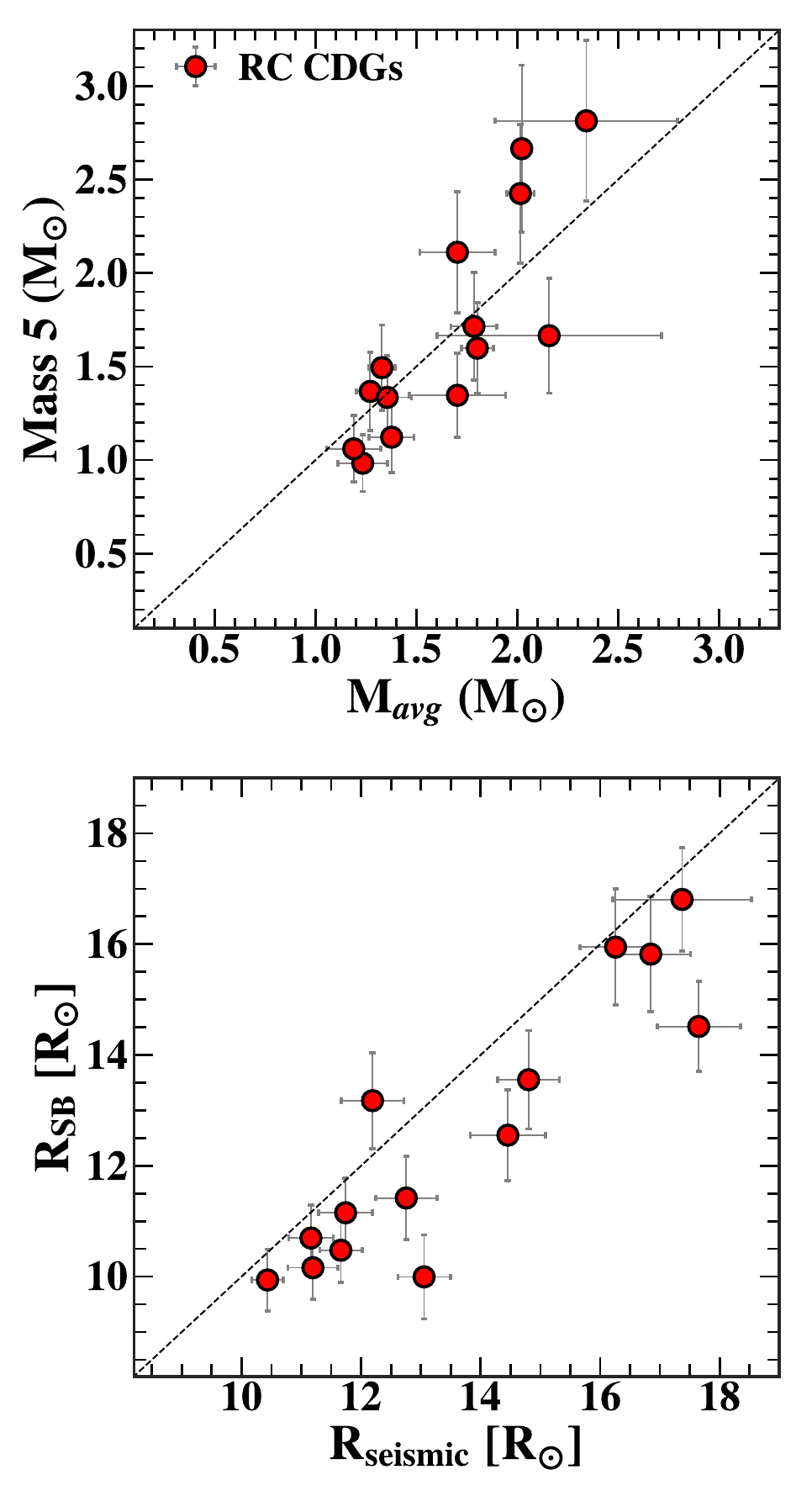}
    \caption{Top panel: Comparison of the average seismic masses of the RC CDGs using Equations (\ref{eq:mass1}), (\ref{eq:mass3}), and (\ref{eq:mass4}) ($\rm{M}_{avg}$; see text for details), versus the non-seismic masses from Equation (\ref{eq:mass5}). The horizontal error bars show the $1\sigma$ standard deviation around the average seismic masses for each star. Bottom panel: Comparison of the seismic radii and radii calculated from the Stefan–Boltzmann law.}
    \label{fig:fig5}
    \end{figure}

    We list our mass determinations for all the 15 CDGs using each of the five mass equations in Table~\ref{tab:table1}. We also show them graphically in Figure~\ref{fig:fig4}, where we compare the individual masses of the RC CDG stars calculated with each equation. 

    As a summary, we show in Table~\ref{tab:table2} the average uncertainties on the masses for each equation, along with the average masses for each equation. It can be seen that the masses calculated with asteroseismic Equation~(\ref{eq:mass2}) have the largest uncertainties ($\overline{err}$ = 1.26~M$_{\odot}$; also see error bars in Figure~\ref{fig:fig4}), when compared to the uncertainties from asteroseismic equations (\ref{eq:mass1}), (\ref{eq:mass3}), and (\ref{eq:mass4}), which all have similar uncertainties ($0.14~\rm{M}_{\odot} - 0.17~\rm{M}_{\odot})$. The non-seismic Equation~\ref{eq:mass5} has larger uncertainties ($0.27~\rm{M}_{\odot}$) than the asteroseismic equations (apart from Eqn.~\ref{eq:mass2}), but is still of the same order of magnitude as the more accurate seismic equations. From now on we do not use the Equation~(\ref{eq:mass2}) masses, due to the large uncertainties.

    For our final seismic masses, we take an average of the three remaining seismic mass equations for each individual star ($M_{avg}$ in Table~\ref{tab:table1}). In Figure~\ref{fig:fig5} we show these masses  versus the non-seismic mass determinations. The seismic mass range of our RC CDGs is 1.2~M$_\odot \leq M \leq 2.3~\rm{M}_\odot$ (with average standard deviation between mass equations of $0.2~\rm{M}_{\odot}$). The non-seismic mass equation gives a slightly expanded mass range of 1.0~M$_\odot \leq M \leq 2.8~\rm{M}_\odot$, with slightly higher average uncertainties ($0.3~\rm{M}_{\odot}$). 

    For the single RGB star, the seismic mass is $1.5~\pm~0.1~\rm{M}_\odot$ and the non-seismic mass is $1.7 \pm \rm{0.3}~M_\odot$.

    \begin{center}
    \begin{table}[ht]
    \centering
    \caption{Average uncertainties on the masses for each mass equation ($\overline{err}$; Eqn.~(\ref{eq:mass1})–(\ref{eq:mass5})). Also shown are the average of the masses given by each mass equation  ($\overline{M}$). It can be seen that Eqn.~(\ref{eq:mass2}) has very large uncertainties. Units are M$_{\odot}$.}
    \label{tab:table2}
    \hspace{-2cm}
    \resizebox{0.8\columnwidth}{!}{%
    \begin{tabular}{ccc}
    \hline
    Mass Eqn.  & $\overline{err} \pm 1\sigma$ & $\overline{M} \pm 1\sigma$  \\
    \hline
    \hline
    1              &   $0.17 \pm 0.08$ & $1.77 \pm 0.46$  \\
    2              & $1.26 \pm 0.77$ & $1.39 \pm 0.29 $      \\
    3              & $0.17 \pm 0.04$   & $1.50 \pm 0.29 $   \\
    4              &  $0.14 \pm 0.06$   & $1.68 \pm 0.37 $  \\
    \hline
    5 (non-seis.)         &  $0.27 \pm 0.09 $  & $1.69 \pm 0.55 $  \\
    \hline
    \end{tabular}%
    }
    \end{table}
    \end{center}

\subsection{Radius determination} \label{sec:radius}

    We calculated the seismic stellar radius for our CDGs using the equation (\citealt{Ulrich1986}; \citealt{Kjeldsen1995}):

    \begin{linenomath}
    \begin{align}
    \left( \frac{R}{R_\odot}\right)  & \simeq \left( \frac{\nu_{max}}{\nu_{max,\odot}}\right) \left( \frac{\Delta \nu}{\Delta \nu_{\odot}}\right)^{-2} \left( \frac{T_{\rm eff}}{T_{\rm eff,\odot}}\right)^{1/2} \label{eq:radius}
    \end{align} 
    \end{linenomath}

    We have used corrected $\Delta \nu$ values of the CDGs, along with the $\nu_{max}$ and $T_{\rm eff}$ values as stated in Table~\ref{tab:table1}. The seismic radii estimates for all 15 CDGs are also provided in Table~\ref{tab:table1}. 
    
    We compare the seismic radius to independent radius estimates calculated using the Stefan–Boltzmann law in Figure~\ref{fig:fig5}. While having greater average uncertainties ($\simeq 0.8~\rm{R}_{\odot}$) than the asteroseismic equation ($\simeq 0.5~\rm{R}_{\odot}$), the non-seismic radius equation is nonetheless of the same order of magnitude as the more precise seismic equation. Between these two methods, all of the RC CDG radii, except two, agree within the $2 \sigma$ uncertainties.

    The seismic radius range of all the RC CDGs is about $10 - 18~\rm{R}_\odot$. Ignoring the two outliers, there is a slight offset between the seismic and non-seismic radii estimates of $\simeq 0.5~\rm{R}_\odot$, where the seismic radii are larger.

    For the RGB star, the seismic radius ($4.6~\pm~0.1~\rm{R}_\odot$) and the non-seismic radius ($4.6 \pm \rm{0.3}~R_\odot$) agree with each other perfectly.

\subsection{Kinematics} \label{sec:kinematics}

    Here we compile the kinematic properties of our sample of CDGs and determine what component(s) of the Galaxy they belong to.
    
    We calculated the Galactic space velocity components (U, V, W) for each star using astrometry with distances taken from the catalog of \citet{BailerJones2021} and proper motions from Gaia~DR3 \citep{Gaia2016, Gaia2022}. The radial velocity data was taken from APOGEE~DR17. We have adopted the Sun's distance from the Galactic center as $8.2$~kpc \citep{Bland-Hawthorn2016}, and its distance above the Galactic plane as $25$~pc \citep{Juric2008}. The space velocity components were computed using the Astropy \texttt{Galactocentric} package \citep{Astropy2013,Astropy2018}. Space velocities are converted to the local standard of rest (LSR) frame using the solar motion ($U_{\odot}$, $V_{\odot}$, $W_{\odot}$)= (11.1, 12.24, 7.25)~km$s^{-1}$ \citep{Bland-Hawthorn2016}. We adopt a Local Standard of Rest velocity $V_{LSR}$ = 232.8~\,km\,s$^{-1}$.

    \begin{figure}[htp]
    \centering
    \includegraphics[width=1\linewidth]{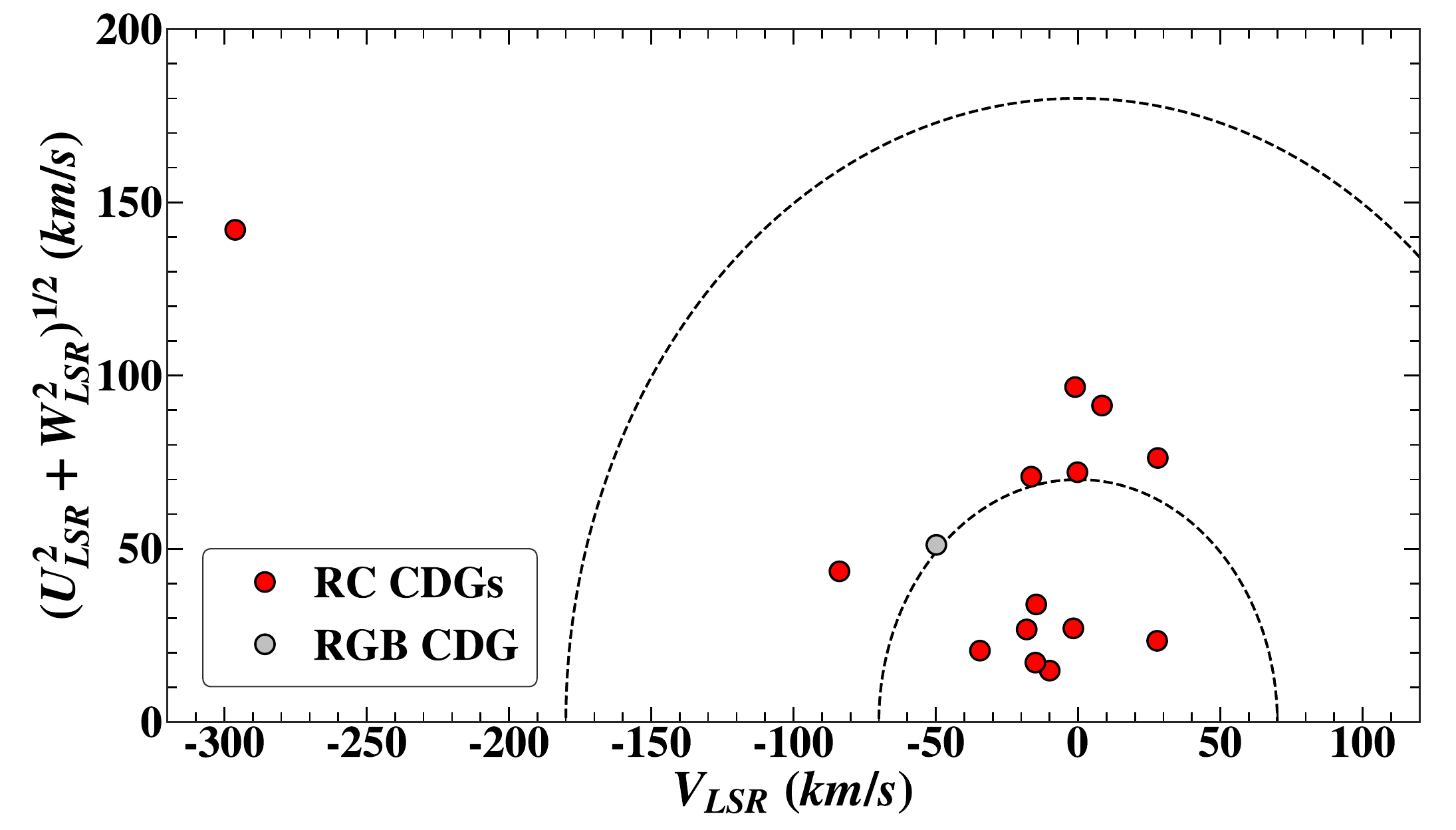}
    \caption{Our CDG sample is shown in the Toomre diagram. Dashed lines show constant values of the total space velocity, $v_{tot} = (U^{2}_{LSR}$+$V^{2}_{LSR}$+$W^{2}_{LSR}$)$^{1/2}$ at 70 and 180\,km\,s$^{-1}$ demarcating thin disk and thick disk components, respectively \citep{Nissen04, Venn2004}.}
    \label{fig:fig6}
    \end{figure}
      
    In Figure~\ref{fig:fig6}, our CDG sample is plotted in the Toomre diagram. We apply limits from \cite{Nissen04} (also \citealt{Venn2004})  that define the different Galactic components. We find our CDGs are distributed in all components of the Galaxy -- the thin disk, thick disc, and one star in the halo.

\subsection{Abundances}
\label{sec:abundances}
\subsubsection{Carbon, Nitrogen and Oxygen} \label{sec:cno_result_section}

\begin{figure*}[ht!]
    \centering
    \includegraphics[width=1.\linewidth]{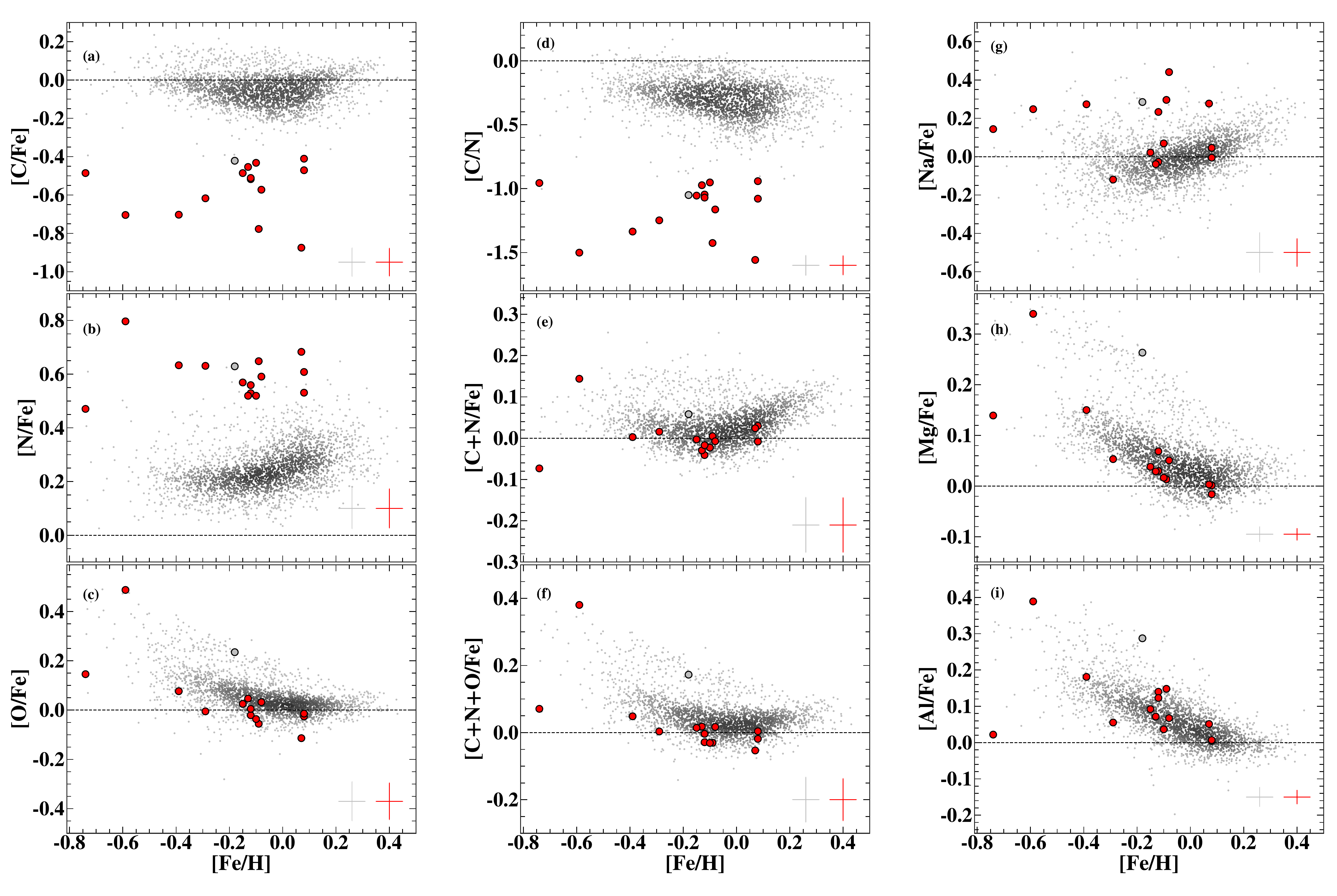} 
    \caption{Various CNO abundance ratios and sums along with sodium, magnesium, and aluminium abundances against [Fe/H] for our CDGs (large symbols). A large sample of RC stars from \cite{Vrard2016} form the background sample (small circles). All abundances are from APOGEE DR17. Average error bars are shown on the lower right of each subplot. Solar abundances are from \citet{Grevesse2007}.} 
    \label{fig:fig7}
    \end{figure*}

    In Figure~\ref{fig:fig7} we plot the C, N, O abundances, the [C+N/Fe] and [C+N+O/Fe] sums, and the [C/N] ratio against [Fe/H] for a background sample of RC stars \citep{Vrard2016}, along with our CDG stars. All abundances are from the APOGEE catalog (DR17; \citealt{Abdurrouf2022}; \citealt{Garcia2016}). All stars have good quality abundances (\texttt{X\_FE\_FLAG=0}; \citealt{Jonsson2020}). The background sample has been limited in mass 1.0~M$_{\odot}$ $\leq M \leq$ 2.5~M$_{\odot}$, to match the mass range of our RC CDGs (Section~\ref{sec:masses}).
    
    The nitrogen abundances of the CDGs are enhanced by an average of $+0.35$~dex over the average [N/Fe] of the background RC sample. This is a smaller offset than for [C/Fe] ($-0.52$~dex). Due to the C and N offsets being in different directions, the average [C/N] offset is large, at $+0.86$~dex. In contrast, the [O/Fe] values of the CDGs generally track the abundances of the background sample. Most are close to scaled solar, except for three stars which have higher [O/Fe]. Two of these stars are our most metal-poor objects. Looking at Mg (Figure~\ref{fig:fig7}), it shows the same pattern as oxygen, so we suggest that these three stars were likely $\alpha$-enhanced from birth.
    
    When we sum C and N we see that most of the stars are scaled solar. The exceptions are the three  $\alpha$-enhanced stars. This again suggests that these 3 stars had different (non-scaled-solar) initial abundances. This is even more evident in the [C+N+O/Fe] plot. 
    
    The strong N over-abundance anti-correlated with the C under-abundance, along with the [C+N+O/Fe] sum being mostly scaled solar, indicates that the material in the atmospheres of CDGs has been processed through CNO cycle hydrogen burning. We see no evidence for O depletion, and therefore infer that the ON cycle was not operating significantly, i.e., only CN cycling occurred.  
    
    Since AGB stars increase [C+N+O/Fe] through third dredge-up (e.g., \citealt{Karakas2014}), they are unlikely to be polluters for our sample. More generally, any scenario in which He-burning products are mixed into the polluting material are not supported by our observations. This is discussed further in Section~\ref{sec:characterisation_of_cdgs}.

\subsubsection{Sodium, Magnesium and Aluminium}

    The Na, Mg, and Al abundances vs metallicity of our CDG stars is shown in Figure~\ref{fig:fig7}, against a background sample of RC stars from \cite{Vrard2016}. 

    Interestingly, Figure~\ref{fig:fig7} shows that the sample splits into two groups in [Na/Fe], one with enhanced [Na/Fe] and one with scaled-solar Na. Taking a cut in [Na/Fe] at $\sim +0.10$~dex, for the Na-enhanced group we find an average sodium abundance of [Na/Fe$] = 0.3 \pm 0.1$~dex, and the Na-normal group has [Na/Fe$] = 0.0\pm 0.1$~dex. We discuss the implications of this Na bimodality (and its correlation with the luminosity bimodality; Section~\ref{sec:bimodality}) in Section~\ref{sec:discussion}.     
    
    The [Mg/Fe] and [Al/Fe] abundances of the CDGs are consistent with those of the background sample. As mentioned, the three stars that appeared to be $\alpha$-enhanced in the [O/Fe]-[Fe/H] plane of Figure~\ref{fig:fig7}, are the ones enhanced with the $\alpha$-element Mg as well.  Thus it is likely that these 3 stars had different (non-scaled-solar) initial abundances, as suggested in Section \ref{sec:cno_result_section}.

\subsubsection{Lithium} \label{sec:lithium_result_section}

\begin{figure}[ht!]
    \centering
    \includegraphics[width=1\linewidth]{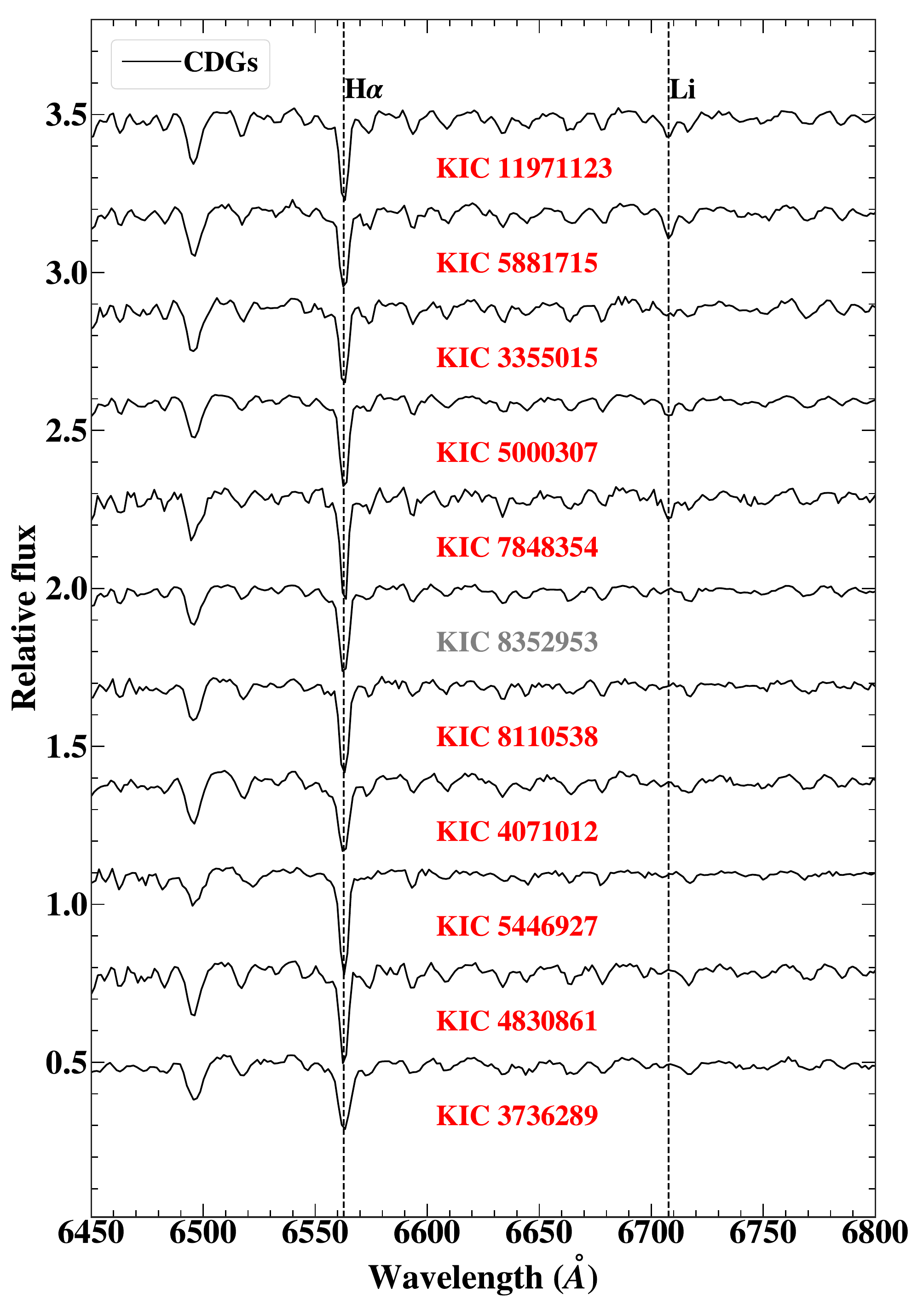} 
    \bigskip
    \includegraphics[width=1\linewidth]{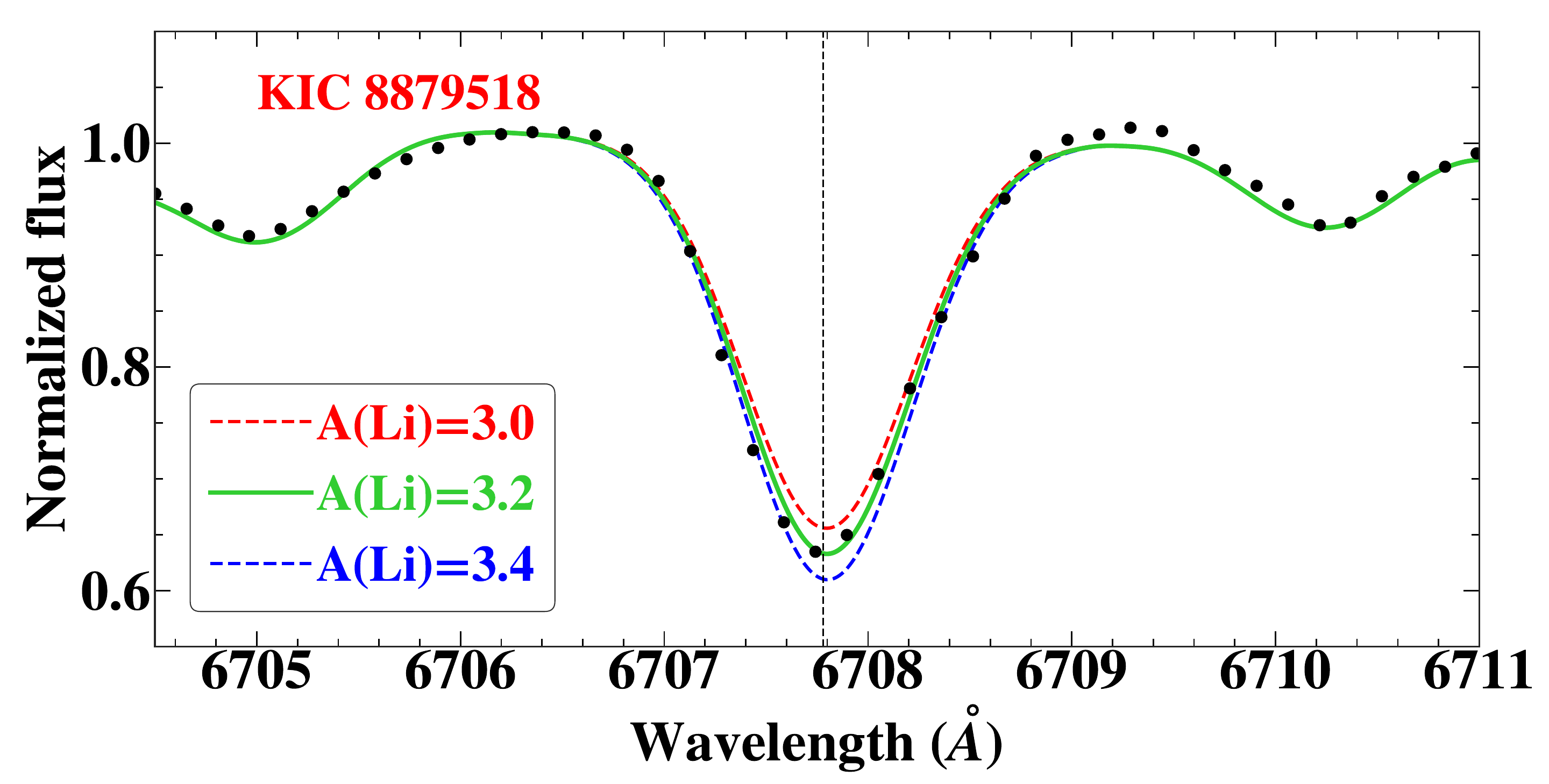}
    \caption{Top panel: LAMOST low-resolution spectra of the CDGs. Note that KIC~8352953 is the only RGB CDG in our sample. Bottom panel: Comparison of the observed LAMOST medium-resolution spectra and synthetic spectra of KIC~8879518 for different Li abundances. Synthetic spectra (green solid line) of A(Li)~=~3.2~dex best fits the observed spectra. The first five CDGs in the top panel and the CDG in the bottom panel show strong Li absorption lines at 6707.8~\AA.}
    \label{fig:fig8}
    \end{figure}

    We cross-matched our CDG sample with the LAMOST survey (\citealt{Zhao2006,Zhao2012}), which contains the $6707.8~\AA$~Li line. We found 11 out of the 15 CDGs to have low-resolution spectra (R $\approx$ 1800; see top panel of Figure~\ref{fig:fig8}). Only one star had a medium-resolution spectrum (R $\approx$ 7500; see bottom panel of Figure~\ref{fig:fig8}). This star (KIC~8879518) has been previously identified to be a super-Li-rich star with A(Li)\footnote{A(Li) = log~($n$(Li)/$n$(H)) + 12, where $n$ is the number density of atoms.}~$= 3.51\pm 0.12$~dex using this spectrum \citep{Raghu2021}.
    
    We determine the Li abundances for the 12 stars by matching synthetic spectra with the observed 6707.8~$\AA$ line. First, the radial-velocity corrected spectra were continuum-fitted and normalized using the iSpec code (\citealt{Blanco2014};  \citealt{Blanco2019}).
    Stellar parameters ($T_{\rm eff}$, $log g, [\rm{Fe/H}]$) were taken from the LAMOST catalog for stars with low-resolution spectra and from the APOGEE catalog for the star with medium-resolution spectrum. The microturbulent velocities ($\xi$) were derived using an empirical relation for giants from \cite{Holtzman2018}. We compiled a line list with associated atomic and molecular data around the Li line using the $linemake$ code\footnote{\url{https://github.com/vmplacco/linemake}} \citep{Placco2021}. 
    
    Local thermodynamic equilibrium (LTE) model atmospheres were generated using the ATLAS9 code \citep{Castelli2003} for the adopted atmospheric parameters. A series of synthetic spectra were then generated for each star by varying the Li abundance, using the Python wrapper of the 2013 version of local thermal equilibrium (LTE) radiative transfer code MOOG \citep{Sneden_thesis}, pyMOOGi\footnote{\url{https://github.com/madamow/pymoogi}}. Finally, the synthetic spectra were matched with the observed spectra. The Li abundance of the best-matched (least $\chi$-square) synthetic spectrum was taken as the LTE abundance for each star. The NLTE Li abundance was then computed following \cite{Lind2009} \footnote{Data obtained from the INSPECT database, version 1.0 (www.inspect-stars.net)}.

    We show the derived A(Li)$_{\rm LAMOST}$ of the CDGs using the LAMOST spectra in Table~\ref{tab:table3}. In the Table, we also present the comparison between A(Li)$_{\rm LAMOST}$ and A(Li)$_{\rm Literature}$ of the CDGs. We see a good consistency between our measurements and the results derived in the literature.

    \begin{table}[]
    \caption{Derived NLTE lithium abundances for our CDGs from LAMOST spectra. Typical uncertainties are $\pm 0.2$~dex.}
    \label{tab:table3}
    \resizebox{0.8\columnwidth}{!}{%
    \begin{tabular}{rcl}
    \hline
    KIC      & A(Li)$_{\rm Literature}$ & ${\rm A(Li)_{LAMOST}}^{1}$ \\
             & (dex)          & (dex)          \\
    \hline
    \hline
    5000307  & 2.7$^{3}$                   & 2.6          \\
    5881715  & 3.4$\pm$0.1$^{4}$           & 3.4          \\
    7848354  & 3.7$^{4}$                   & 3.3          \\
    11971123 & 2.9$^{4}$                   & 2.9          \\
    3355015  & 1.9$^{4}$                   & 2.1          \\
    5446927  &                             & $<$1.8       \\
    4071012  &                             & $<$1.8       \\
    3736289  &                             & $<$1.8       \\
    8110538  &                             & $<$1.8       \\
    8352953  &                             & $<$1.8       \\
    4830861  &                             & $<$1.8       \\
    \hline
    8879518$^{2}$  & 3.5$\pm$0.1$^{5}$    & 3.2         \\
    \hline
    \end{tabular}%
    }
    \begin{quoting}
          {\small {\bf Note:} $^{1}$This work;
          $^{2}$Only has LAMOST medium-resolution spectrum;
          $^{3}$\cite{Silva2014};
          $^{4}$\cite{Yan2021};
          $^{5}$\cite{Raghu2021}.
     }
    \end{quoting}
    \end{table}

    In total we found 6 out of the 12 CDGs with LAMOST spectra to be Li-rich. These stars are also identified as Li-rich in the \cite{Yan2021} study (Table~\ref{tab:table3}). Thus 50\% of our sample for which we have spectra is Li-rich. This can be compared to the expected fraction for field stars. First we note that our detection limit, A(Li)~$> 1.8$~dex, is similar to the traditional Li-rich definition, A(Li)~$> 1.5$~dex. Using a GALAH survey sample, \citealt{Kumar2020} showed that the fraction of RC stars with  A(Li)~$> 1.5$~dex is $\approx 3\%$. Thus our finding of $50\%$ represents an extremely high fraction -- $17$ times the expected fraction -- despite the selection criterion being on carbon, not lithium. Further, for RC stars it has been shown that stars with A(Li)~$> -0.9 $~dex can be considered Li-\textit{enhanced}, due to Li destruction in the preceding RGB phase (at least for stars in the mass and metallicity range considered in \citealt{Kumar2020}). Thus many more CDGs in our sample may be Li-enhanced if studied with high-resolution spectra -- our 50\% is a lower limit in this regard. Regardless of the exact proportion, it is clear that Li-richness is highly correlated with C-deficiency. We discuss this further in Section~\ref{sec:link_to_lithium}.

\subsection{Bimodality of the RC CDGs}  \label{sec:bimodality}
\subsubsection{Over-luminous and normal-luminosity stars} \label{sec:luminosity_bimodality_section}

   In Section~\ref{sec:luminosity} we reported a bimodality in the luminosity distribution of our RC CDGs (see Figure~\ref{fig:fig3}). The fainter CDGs are in a region bounded by $\log (L/L_{\odot})$ from $1.67$~dex to $1.89$~dex, which matches well with theoretical predictions of the core He-burning phase for low-mass stars (i.e., $\log (L/L_{\odot}) \simeq 1.55- 1.85$~dex; \citealt{Girardi2016}). This is also clear from Figure~\ref{fig:fig11}a where we compare our RC CDG sample with a background RC sample -- the fainter RC CDGs fall on the normal RC. We will be referring to these stars as `Normal-luminosity-RC CDGs'.

   The brighter CDGs are more luminous than the typical RC stars, they have $\log (L/L_{\odot})$ between $1.98$~dex and $2.17$~dex. RC models of masses covering our seismic mass range do not have luminosities this high. Observationally, we see that the bright CDGs are a factor of $\approx 2$ brighter than the normal RC for stars of the same mass ($\approx 60~\rm{L}_\odot$ vs $120~\rm{L}_\odot$; Figure~\ref{fig:fig11}a). We refer to this group as `Over-luminous-RC CDGs'.
    
    It is possible that the bimodality in the luminosity distribution is an artefact of systematics in our luminosity determination. In Figures~\ref{fig:fig10}c and \ref{fig:fig10}d we show the asteroseismic parameters for our sample. These show correlations with luminosity, with the over-luminous stars having low $\Delta\nu$ and $\nu_{max}$. Since luminosity and the seismic parameters are totally independent measurements, this strongly suggests that the bimodality is real. Moreover, in Figure~\ref{fig:fig10}h we see a very strong correlation with sodium, with the over-luminous stars generally having higher Na than the normal-luminosity-RC CDG stars, by an average of $+0.25$~dex. This further supports that these are two distinct groups of RC stars.
    
    Radius shows an expected correlation with luminosity. On the other hand, mass does not show a significant correlation -- both groups roughly cover the same mass range ($\simeq 1.2$ to 2.3~M$_{\odot}$), although the normal-luminosity group appears biased to slightly lower masses on average, with $M_{avg} = 1.5~\rm M_{\odot}$ compared to $1.8~\rm M_{\odot}$ for the over-luminous stars. 
    
    In the bottom panel of Figure~\ref{fig:fig10} we also show the C, N, and O abundances of our sample versus luminosity. The over-luminous stars have slightly lower carbon than the normal-luminosity stars. We find an average carbon abundance of [C/Fe$] = -0.7 \pm 0.1$~dex for the former and [C/Fe$] = -0.5\pm 0.1$~dex for the latter. The nitrogen abundances of the over-luminous stars are enhanced by an average of $+0.1$~dex more than the average [N/Fe] of the normal-luminosity CDG stars. This is a smaller offset than the average for carbon ($-0.2$~dex). In contrast, the [O/Fe] values of both groups of stars are close to scaled solar, although 3 over-luminous stars appear to be $\alpha$-enhanced (Section~\ref{sec:abundances}).
    
    \subsubsection{Known-CDGs in context of luminosity bimodality} \label{sec:known_cdgs_bimodality_section}

    In the bottom panel of Figure~\ref{fig:fig10} we also show 29 known-wGb stars from the literature that have carbon abundances from high-resolution optical spectra \citep{Adamczak2013, Palacios2016}. Of these, only 6 stars have $\nu_{max}$ reported in the literature, based on TESS data \citep{Hon2021}. These 6 stars are plotted in the top panel of Figure~\ref{fig:fig10}. It can be seen that these known-wGb stars have high luminosities and low $\nu_{max}$, similar to our sample of over-luminous stars. They also have similar $T_{\rm eff}$ to our luminous sample, and therefore lie in the same region of the HRD in Figure~\ref{fig:fig11}a. This suggests that they are also RC stars, and appear to be the same type of stars as our over-luminous-RC CDGs, just with more extreme chemical signatures (Figure~\ref{fig:fig10}).

    However, despite being over-luminous, this literature sample have spectroscopic log(g) values close to the normal-luminosity stars (average log(g)~$\simeq 2.7 \pm 0.1$~dex; also see Figure~\ref{fig:fig1}), and thus appear to be systematically high in log(g) given their luminosities. We note that the log(g) values are from the literature, so may have a systematic offset from our sample. No seismic log(g) is available for this known-wGb sample as a cross-check. More data is required to determine the source of this discrepancy. In contrast, the luminosities of these stars are on the same scale as our sample.  
    
    Combining $\nu_{max}$ with the non-seismic parameters ($T_{\rm eff}$, $L/L_{\odot}$), we estimated the seismic masses of these 6 stars using mass Equation~\ref{eq:mass3}. The seismic mass range of these known-wGb stars is 1.6~M$_\odot \leq M \leq 2.9~\rm{M}_\odot$ (with average $1\sigma$ uncertainties on the masses of $0.3~\rm{M}_\odot$). With the exception of one star at $\simeq 3~\rm{M}_{\odot}$, the rest of the sample is consistent with being low-mass (M~$\leq 2~\rm{M}_{\odot}$; Figure~\ref{fig:fig10}b), similar to our Kepler-APOGEE sample.
    This is in contrast to previous studies which have reported that these 6 known-wGb stars are intermediate-mass stars ranging from 2.9~$\rm M_{\odot}$ to 3.3~$\rm M_{\odot}$, based on their positions on the HRD (\citealt{Adamczak2013}; \citealt{Palacios2016}). 
    
\begin{figure}[htp]
    \centering
    \includegraphics[width=0.9\linewidth]{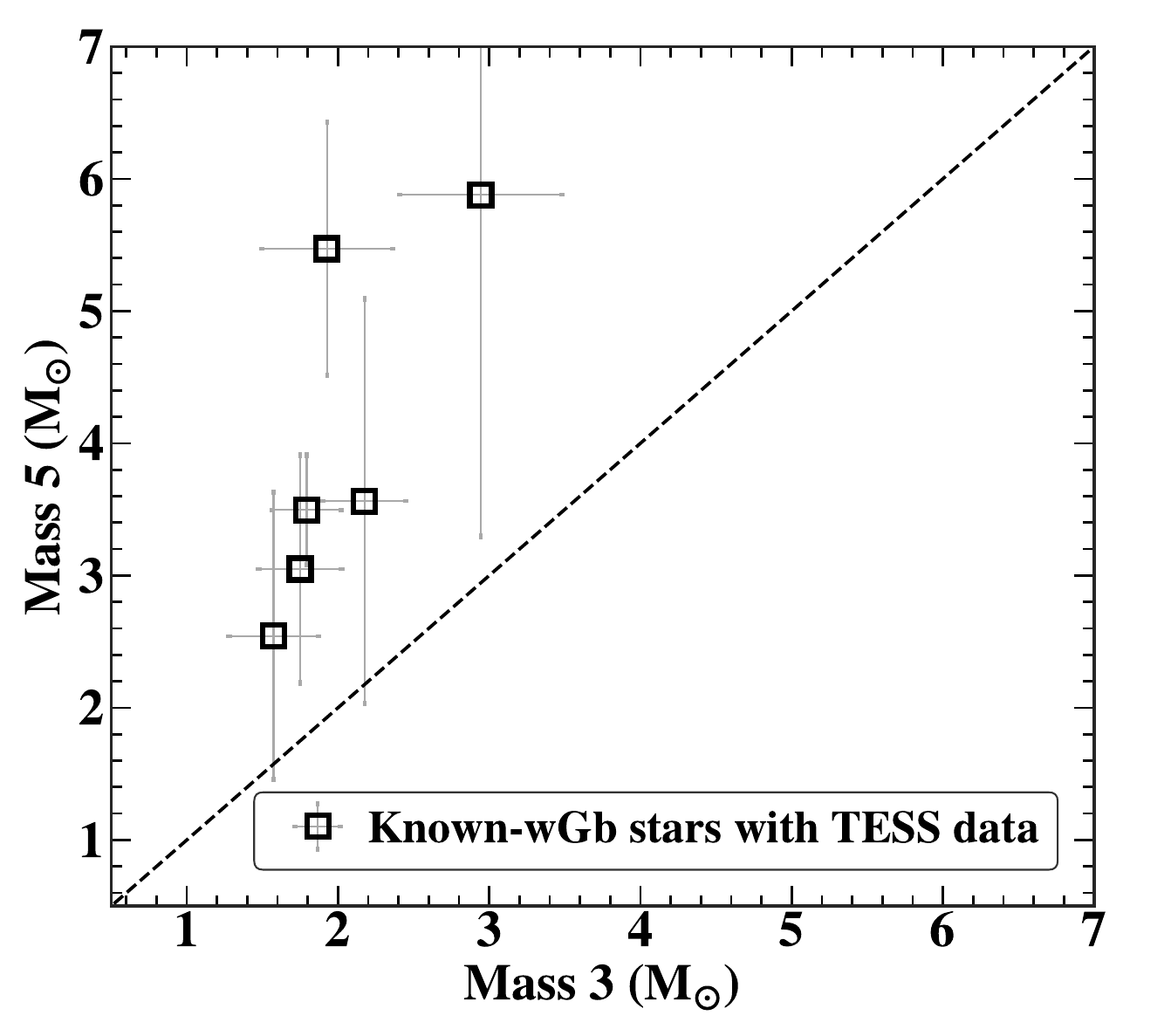}
    \caption{Comparison of the seismic masses of the previously-known-wGb stars with TESS data determined using Equation~(\ref{eq:mass3}), versus their masses determined using the non-seismic mass Equation~(\ref{eq:mass5}).}
    \label{fig:fig9}
    \end{figure}

    As a check, we also calculated the non-seismic masses (mass Equation~\ref{eq:mass5}) for this sample. We found that these masses were substantially higher. In Figure~\ref{fig:fig9} we compare these two mass estimates. We are unsure as to why there are such large differences, but it may be related to the log(g)-luminosity tension noted above. Further, the uncertainties on the non-seismic masses are generally very large. In fact, 4 out of the 6 stars have non-seismic masses within 1.5$\sigma$ of the seismically-determined masses (Figure~\ref{fig:fig9}). We note that this discrepancy is not found for our Kepler-APOGEE sample, where the seismic masses are consistent with the non-seismic masses (Figure~\ref{fig:fig5}). 
    
    In terms of chemical abundances, the 6 known-wGb stars have more extreme chemical patterns than our sample of stars. Their C is more depleted and N more enhanced. Sodium, which is scaled solar in the normal-luminosity group and enhanced by an average of [Na/Fe$] = +0.2$~dex in our over-luminous group, is further enhanced in the known-wGb sample ([Na/Fe$] = +0.4$~dex). Oxygen, in contrast with our sample which is approximately scaled solar, is on average slightly enhanced, with [O/Fe$]\sim +0.1$~dex. We are unsure whether this is due to their initial composition being $\alpha$-enhanced due to their lower average [Fe/H], or if it may indicate a further nucleosynthetic product of the pollution episode. This is explored further in the Discussion.

 \subsubsection{Seismic data and the luminosity bimodality: Masses and Radii} \label{sec:olrd_nlrc_seismic_data_section}

\begin{figure*}[ht!]
       \centering
    \includegraphics[width=\linewidth]{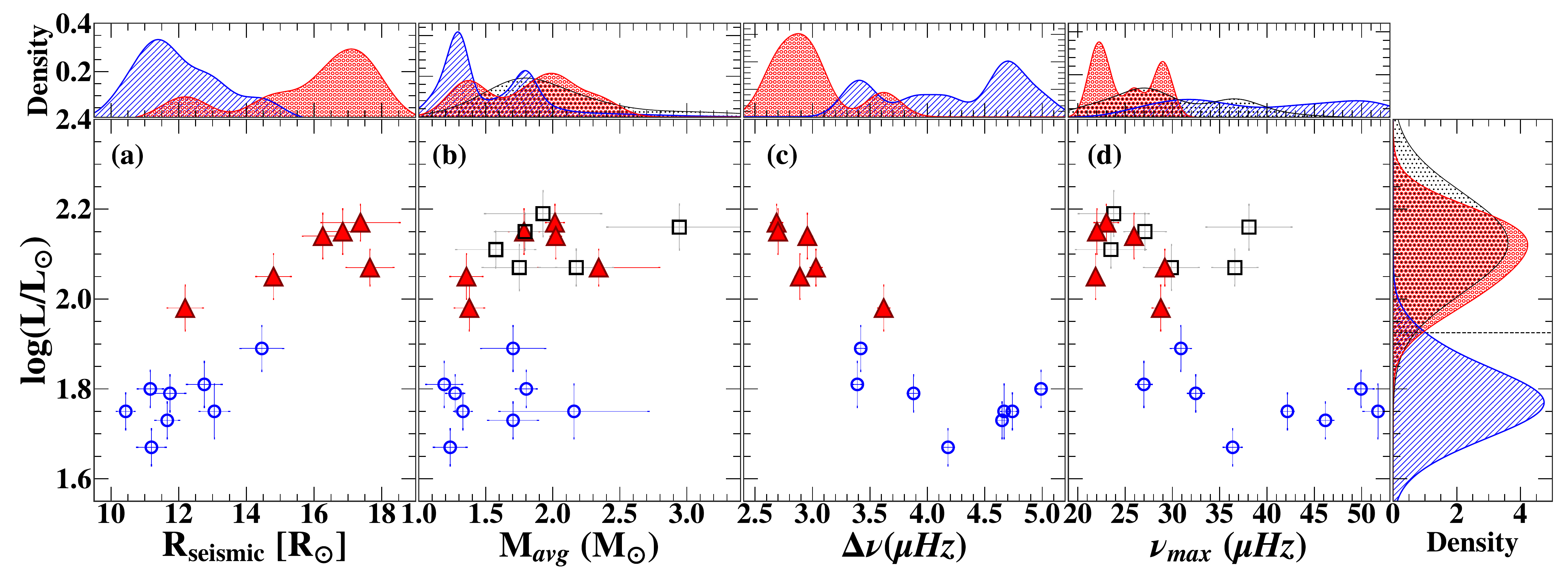}
    \bigskip
    \includegraphics[width=\linewidth]{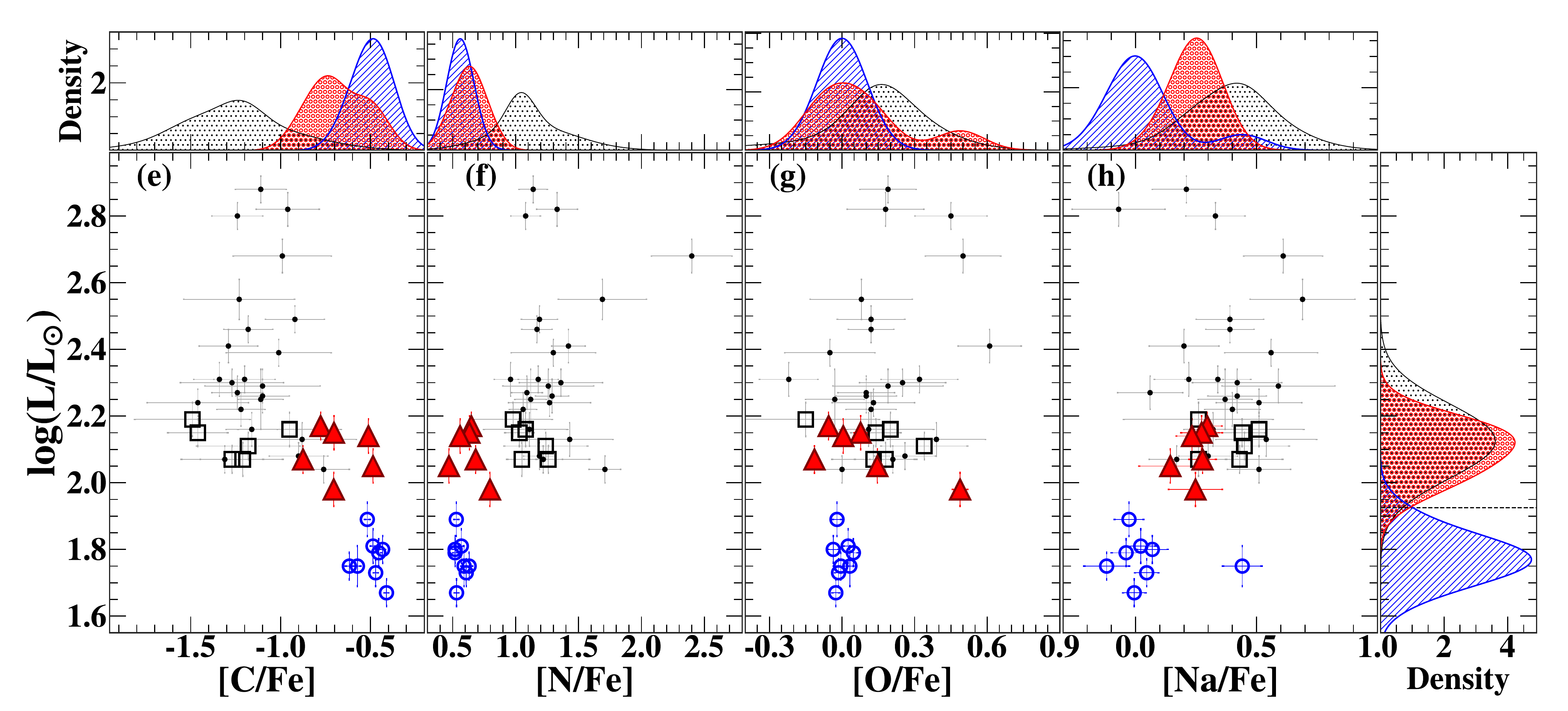}
        \caption{Top panel: $\log(L/L_{\odot})$ as a function of asteroseismic radius, mass, $\Delta \nu$, and $\nu_{max}$. Bottom panel: $\log(L/L_{\odot})$ as a function of the carbon, nitrogen, oxygen, and sodium abundances. Kernel density histograms are included to show the distributions for the RC CDGs, with the color scheme following the sample. The horizontal dashed line at $\log(L/L_{\odot})~= 1.925$~dex, denotes the cut we use to define two different luminosity groups of our RC CDG sample (see text for details). The large filled red triangles and open blue circles show our over-luminous-RC CDGs and normal-luminosity-RC CDGs, respectively. Large black squares are the known-wGb stars that have asteroseismic parameters from \cite{Hon2021} and the small black circles are the rest of the known-wGb stars from the literature. The abundances of the known-wGb stars are from high-resolution optical spectra \citep{Adamczak2013, Palacios2016}. In order to compare the abundances of the known-wGb with that of our CDG sample, they need to be on the same scale. We apply the average offset of $+0.02$, $-0.15$, $-0.12$, $-0.11$, and $-0.03$~dex to the [C/Fe], [N/Fe], [O/Fe], [Na/Fe], and [Fe/H] values of known-wGb respectively, to  account for the systematic difference between the abundances from IR and optical spectra \citep{Jonsson2020}. The top and right panels of all the subplots show kernel density histograms for the RC CDGs (dashed blue lines and large red dots) and known-wGb stars with asteroseismic parameters (small black dots). We use the solar abundance of C, N, O, and Na as derived by \cite{Grevesse2007}.}
    \label{fig:fig10}
    \end{figure*}

\begin{figure*}[htp]
    \centering
    \includegraphics[scale=3,width=\textwidth]{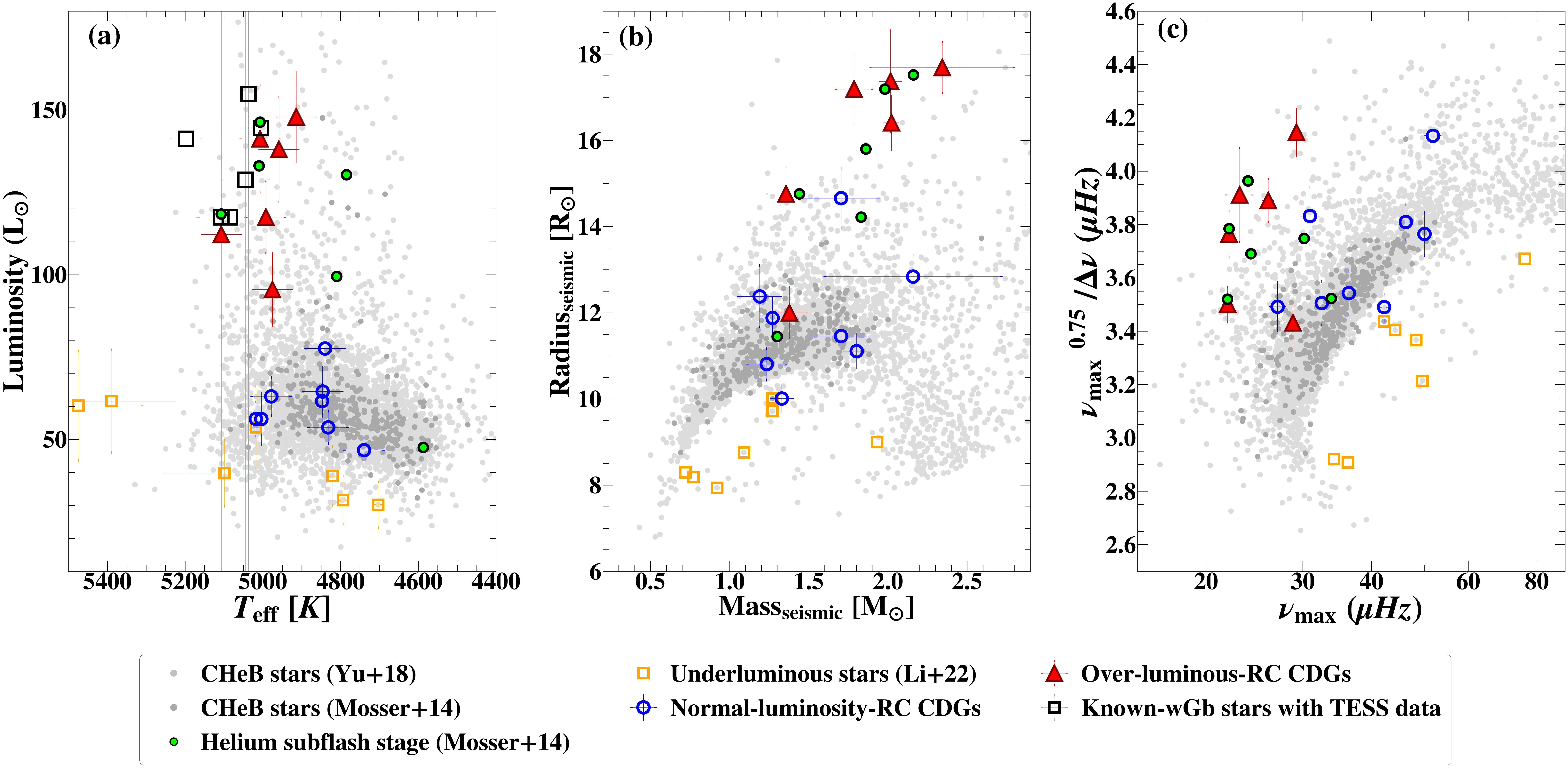} 
    \caption{(a) The Hertzsprung–Russell diagram. (b) The mass-radius 
    diagram. (c) The seismic quantity ${\rm \nu_{max}}^{0.75}/ \Delta \nu$ versus $\rm \nu_{max}$. The RC stars classified based on asteroseismic analysis form the background from \cite{Mosser2014} and \cite{Yu2018} (small filled grey and silver circles respectively). The helium subflash stars from \cite{Mosser2014} are represented by small filled green circles. The underluminous stars from \cite{Li2022} are shown by orange squares, the known-wGb stars by black squares, the normal-luminosity-RC CDGs, and the over-luminous-RC CDGs from our study by blue circles and red triangles respectively.}
    \label{fig:fig11}
    \end{figure*}

    In panels~(b) and (c) of Figure~\ref{fig:fig11} we plot seismically-derived parameters, following the analysis of \cite{Li2022} into underluminous and stripped stars. We note that in all panels of Figure~\ref{fig:fig11}, our over-luminous-RC CDGs are concentrated in regions far from the normal (background) RC stars. 
    This is true for panel~(a), which is based on photometric data, and also panel~(c) which is entirely based on seismic data. As mentioned above, the fact that the over-luminous-RC CDGs stand out in diagrams with totally independent data (and in chemistry; see Figure~\ref{fig:fig10}) is a very strong indication that the bimodality is real.

    The reason the over-luminous-RC CDGs stand out in panel~(c) is because they have low $\nu_{max}$ and low $\Delta \nu$ for their masses, given their phase of evolution. This suggests that they have different structures compared to normal RC stars in the same mass range. The middle panel of Fig~\ref{fig:fig11} shows that the majority of these stars also have abnormally large radii, with the 4 largest stars (out of 6) having an average radius of $R_{avg} =17.0 \pm \rm{0.8}~\rm{R}_\odot$ compared to the background sample with $R_{avg} =11.5~R_\odot$ in the mass range of interest. Thus there is a radius difference of $\simeq 5.5~\rm{R}_{\odot}$. Of the other two stars, one has a high radius for its mass (although not as big a difference as the others), and the other has a radius consistent with normal RC stars.

    The normal-luminosity-RC CDGs (8 stars) mostly have normal RC radii ($R_{avg} =11.7 \pm \rm{0.4}~R_\odot$)\footnote{This average ignores the outlier.}, although there is one outlier that has a large radius. Also, one star has slightly lower radius than expected for the RC. This star (KIC~5000307) is classed as an underluminous star (partially stripped) by \cite{Li2022}. Although it does stand out in the seismic diagram we find its luminosity to match that of normal RC stars (Figure~\ref{fig:fig11}a).

    We note that the over-luminous-RC CDGs in all panels of Figure~\ref{fig:fig11} coincide with the position of stars in the helium subflashing phase as defined by \cite{Mosser2014}. We discuss this in Section~\ref{sec:olrc_subflash}.
    
    Finally, as mentioned above, the normal- and over-luminous groups cover roughly the same mass range. However we see in Figure~\ref{fig:fig11}b that both groups have higher average masses than the background RC sample. In particular, there are no RC CDGs below about $1.2~\rm M_{\odot}$. Further, the over-luminous stars appear biased to slightly higher masses ($M_{avg} = 1.8~\rm M_{\odot}$), while the normal-luminosity stars have an average mass of $1.5~\rm M_{\odot}$. Being relatively massive is suggestive of a merger-product population, which we discuss in detail in Sections~\ref{sec:OLRC_merger}~and~\ref{sec:NLRC_merger}.

\section{Discussion} \label{sec:discussion}

    The first dedicated spectroscopic survey for C-deficient stars was undertaken 50 years ago by \cite{Bidelman1973}, who found 34 wGb/CDG stars. This sample has comprised 77\% of the wGb stars studied through the decades (e.g., \citealt{Sneden1978,Cottrell1978,Parthasarathy1980,Palacios2012,Adamczak2013, Palacios2016}). \cite{Bond2019} added 5 more wGb stars identified via spectroscopy. The primary aim of the current study was to better characterize the CDGs since they are still poorly understood.

\subsection{Overview of our results} \label{sec:results_overview}

     To better characterize the CDGs we took advantage of the high-quality asteroseismic data of the Kepler field, which allows mass and evolutionary-phase determinations, combined with data from large spectroscopic surveys that overlap with the Kepler field (APOGEE and LAMOST), which provide abundances and stellar parameters. Here we enumerate our key results, and we discuss the implications of these findings further below:
    
    \begin{enumerate}
    \item To date, there has been no strict definition for CDGs in literature. We developed a definition for CDGs based on stellar model predictions and a large spectroscopic sample. Standard stellar models do not deplete carbon below $\approx -0.3$~dex (taking into account lower than average initial abundances), and we consider stars with [C/Fe]~$< -0.4$~dex as carbon deficient. 
    \item Our sample adds 15 new CDGs to the literature. All of our CDGs are in the Kepler field. They have quality abundances, asteroseismic masses and evolutionary phases. 
    \item Considering our new seismic masses, most of our CDGs are low-mass stars, that is, stars that are expected to have gone through the core helium flash ($M \lesssim 2~\rm{M}_{\odot}$). Taking into account mass uncertainties of the 2 most massive stars in the sample, 100\% of the sample is consistent with being low-mass ($1~\rm M_{\odot} - 2~\rm M_{\odot}$). This is at odds with the previous understanding in the literature, where wGb stars were considered to be of intermediate mass ($M = 2.5~\rm M_{\odot} - 5.0~\rm M_{\odot}$), which do not experience the core helium flash. 
    \item Remarkably, we conclude that 93\% of our CDGs are in the RC phase -- 14 out of 15 stars -- with only one being an RGB star. This is strongly supported by our subsample of 11 stars for which we have asymptotic g-mode values $\Delta\Pi_{1}$, $100\%$ of which were found to be RC stars (Figure~\ref{fig:fig2}).   
    \item Two out of the 11 stars (18\%) for which we have $\Delta\Pi_{1}$ measurements are in the helium subflashing phase according to \cite{Mosser2014}. This is 4.5 times the expected percentage (3.8\%) found in our background sample of RC stars. Both stars are over-luminous-RC CDGs. Moreover, in the seismic diagram of Figure~\ref{fig:fig11} almost all of the over-luminous-RC CDGs are found in the same region as the subflashing stars.

    \item We found that the CDGs are universally N-enhanced. Analysis of CNO sums showed that the material in the envelopes of our CDG stars has been processed through the CN cycle, and likely not the ON cycle. The fact that [C+N+O/Fe] is scaled-solar indicates that no He-burning products have been involved in the pollution event(s) (although this is not true for the literature sample in Section~\ref{sec:chemical_signatures}).
    
    \item We find a strong correlation between stars being C-deficient and Li-rich, with 6 out of 12 CDGs (50\%) for which we have low-resolution spectra being Li-rich. This is 17 times higher than the expected fraction of 3\%.    
    \item We find a bimodality (roughly 60:40) in luminosity within our sample, with the normal-luminosity-RC CDGs (60\% of the sample) having an average luminosity of $\log(L/L_{\odot}) \simeq 1.8$~dex ($60~\rm{L}_\odot$) , and 
    the over-luminous-RC-CDGs having an average luminosity of $\log(L/L_{\odot}) \simeq  2.1$~dex ($120~\rm{L}_\odot$). Given their low masses, the over-luminous-RC-CDGs are significantly more luminous than expected, by a factor of about 2.    
    \item On further investigation we found the luminosity groups to also be distinct in Na abundance, with the normal-luminosity-RC CDGs having scaled solar Na but the over-luminous-RC CDGs having on average [Na/Fe]~$\sim 0.3$~dex. This is a factor of about 1.5 greater than expected from FDU. The luminosity groups also strongly  correlate with radius and the asteroseismic parameters, with the more luminous stars having lower $\Delta\nu$ and lower $\nu_{max}$. In addition, the over-luminous-RC-CDGs have lower carbon and higher N than the normal-luminosity-RC CDGs, indicating more extreme pollution.   
    \item Comparing our sample with the previously-known wGb stars for which we have some seismic data (TESS), we found that the latter are also primarily low-mass stars and are over-luminous for their masses. Thus they appear to be members of our over-luminous-RC group. They however show signs of having undergone even more extreme pollution.
    \end{enumerate}

\subsection{Characterisation of carbon-deficient stars} \label{sec:characterisation_of_cdgs}

\subsubsection{Three groups of CDGs}

    As reported in Section~\ref{sec:bimodality} (also see points~9 and 10 above), our CDGs appear to fall into two distinct groups: normal-luminosity-RC and over-luminous-RC stars. Before we discuss the possible formation theories for CDGs, here we explore the differences between these groups further and compare to other, previously-known CDGs in the literature.

    In Figure~\ref{fig:fig12} we show [Na/Fe] versus the sum [C+N+O/Fe] for various stellar samples. We see here that our two luminosity groups are clearly separated in Na. However, they are not separated in [C+N+O/Fe], with both showing scaled-solar composition. By contrast, the known-wGb stars for which we have seismic data (open squares, 6 stars) are enhanced in Na but also in [C+N+O/Fe] (by an average of $+0.2$~dex). A larger literature sample, which has no seismic data (filled circles, 23 stars), is also shown in Figure~\ref{fig:fig12}. These stars overlap with our  sample -- they are also simultaneously Na-rich and CNO-rich. Thus it appears that the CDGs fall into three groups.

    To further investigate these possible groupings, we collate all key information we have from the current study: abundances, masses, and luminosities. The groupings do appear to be distinct -- we display the characteristics for each in Table~\ref{tab:table4}, and include the non-seismic literature sample as Group~3b. In this table, we see a clear progression from lower luminosities to higher luminosities, along with lower masses to higher masses, although there is a significant overlap between the groups. There is also a progression of C-depletion (N-enhancement) through the groups. Oxygen is only enhanced in Group~3. Lithium is anti-correlated with Na in the first two groups, with the normal-luminosity-RC CDGs being Li-rich but the over-luminous-RC CDGs not. At the more massive/luminous end (Group~3), Li-rich stars appear again. 

\subsubsection{Chemical signatures in the groups}
\label{sec:chemical_signatures}

    Having [C+N+O/Fe]~$> 0.0$ is a signature of He-burning products being mixed up to the envelope, whilst [C+N+O/Fe]~$ =0.0$ is a signature of CN(O) cycling only, since in this case the CNO elements are just transmuted within the cycle (primarily to N). Given a scaled-solar starting composition during a CNO burn, there is a limit to which N can be enhanced via the CN cycle. As an estimate, taking the extreme limit where all C is burned to N, we would expect [N/Fe] to reach its upper limit at $\simeq +0.6$~dex. If the ON cycle were activated, higher N abundances would be possible. Given that our sample of CDGs (Groups~1 and 2) generally has scaled-solar oxygen (Figure~\ref{fig:fig7}; Table~\ref{tab:table4}) and [C+N+O/Fe]~$=0.0$, it appears the ON cycle was not activated in these stars. However, due to the very high initial abundance of O, it would take very little ON burning to increase the N. Taking a typical uncertainty on the O abundances as $\simeq 0.1$~dex, a rough calculation shows that a hidden (within uncertainties) 0.1~dex depletion in O would increase [N/Fe] by another 0.2~dex, to [N/Fe]~$=0.8$~dex. We indicate this limit in Figure~\ref{fig:fig13} (vertical shaded band, since it is approximate), which shows the C and N distributions of all the CDG groups. The limit matches well with the observed upper [N/Fe] limit of our CDGs. Any stars with [N/Fe] above this limit must have been enhanced in CNO elements at some stage.
    
\begin{figure*}[htp]
    \centering
    \includegraphics[width=0.9\linewidth]{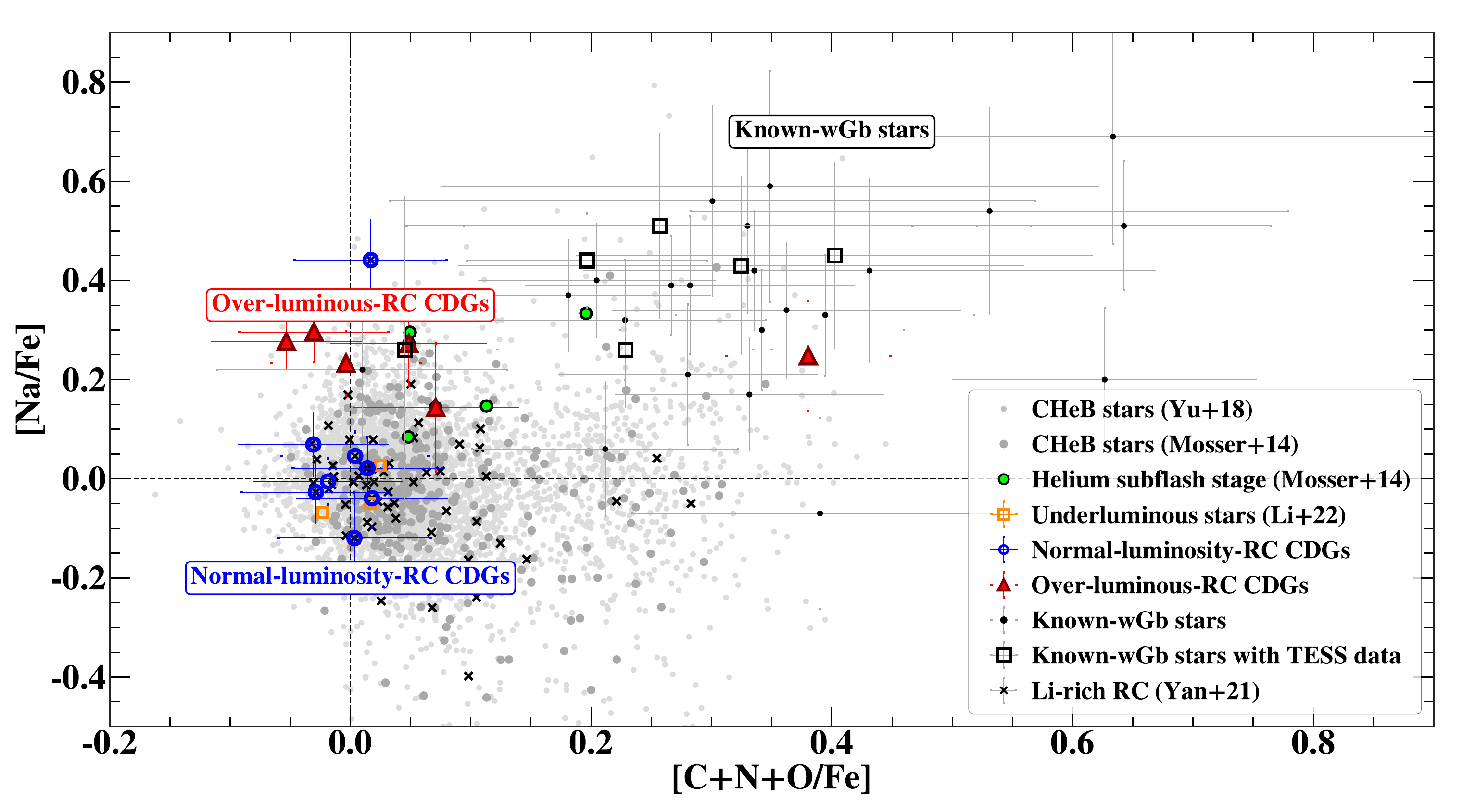} 
    \caption{Trends of sodium abundance vs. the [C+N+O/Fe] abundance ratio. Giants classified based on asteroseismic analysis form the background from \cite{Mosser2014} and \cite{Yu2018} (small filled circles; see key). The underluminous stars from \cite{Li2022} are shown by orange squares. The abundance ratios of the normal giants and the underluminous stars are from APOGEE~DR17. The normal-luminosity-RC CDGs, and the over-luminous-RC CDGs from our study are represented by blue circles and red triangles respectively. Large black squares are the previously-known-wGb stars that have asteroseismic data from \cite{Hon2021} and the small black circles are the rest of the known-wGb stars from literature with carbon abundances from high-resolution spectra. The Li-rich giants (black crosses) are from the \cite{Yan2021} study. These Li-rich giants are the low-spectral-resolution RC sample that have good-quality APOGEE~DR17 data. In total, we have a sample of 59 Li-rich RC stars for comparison with the CDGs. Six of our normal-luminosity-RC CDGs are included in the Li-rich study of \cite{Yan2021}. We use the solar abundance of C, N, O, and Na as derived by \cite{Grevesse2007}.}
    \label{fig:fig12}
    \end{figure*}

\begin{figure*}[htp]
    \centering
    \includegraphics[scale=5,width=\linewidth]{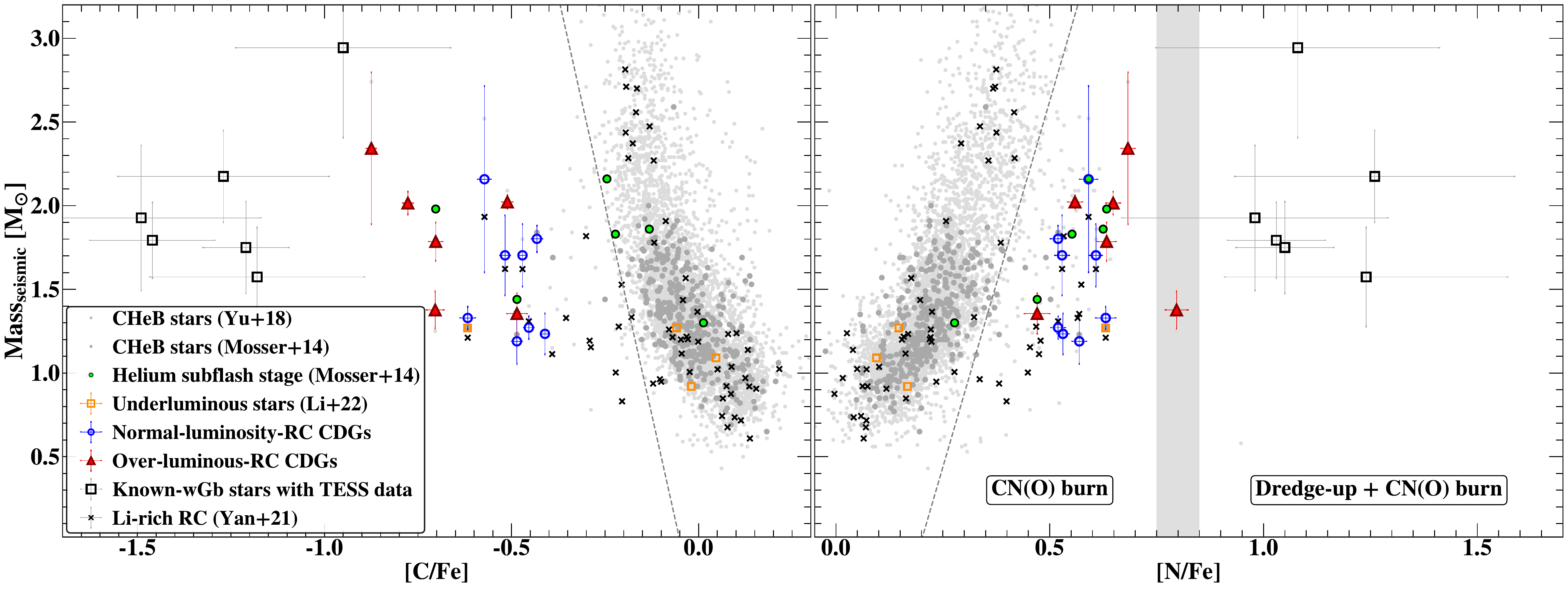} 
    \caption{Seismic mass versus [C/Fe] (left panel); seismic mass versus [N/Fe] (right panel) for our CDGs, the previously-known-wGb stars, underluminous stars from \cite{Li2022}, Li-rich giants from \cite{Yan2021} and a large sample of RC stars from \cite{Mosser2014} and \cite{Yu2018}  with symbols having the same meaning as in Figure~\ref{fig:fig12}. We determine luminosities for the Li-rich giants in the same manner as our CDGs, detailed in Section~\ref{sec:luminosity}. Using this, we estimated the seismic masses of these stars using  mass Equation~\ref{eq:mass3}. Six of our normal-luminosity-RC CDGs are included in the Li-rich study of \cite{Yan2021}. They overlap with the Li-rich giants with a slight offset in seismic mass. The `normal' RC stars show a correlation between mass and nitrogen (and carbon) abundance. This reflects first dredge-up surface pollution from early on the RGB (e.g., \citealt{Iben1984}). The trend shows fairly sharp edges, allowing us to make a cut in N (or C) that varies with mass (dashed lines). This helps in distinguishing the chemically peculiar stars. To the right of the N line the stars can be considered N-rich, for their mass. The vertical shaded region at [N/Fe$] = +0.8\pm0.05$~dex, denotes the upper limit of our CDG sample’s N enhancement as well as the  highest N abundance possible if the ON cycle was activated after all C is burned to N (see text for details).  Solar abundances are from \citet{Grevesse2007}.}
    \label{fig:fig13}
    \end{figure*}

    The previously-known CDGs with TESS seismic data (open squares in Figs.~\ref{fig:fig12} \& \ref{fig:fig13}; Group~3a in Table~\ref{tab:table4}) have much higher N abundances than our Group~1 and 2 stars, starting at [N/Fe]~$\simeq +1.0$~dex and increasing to $\simeq 1.3$~dex. Thus there appears to be a nitrogen gap between Groups~1+2 and Group~3. Given Group~3 is above the [N/Fe] limit of +0.8~dex for complete burning of C to N, these stars must have had CNO elements added to their envelopes, likely through dredge-up of He-burning products. This aligns with their overabundances of [C+N+O/Fe] (Figure~\ref{fig:fig12} and Table~\ref{tab:table4}). However, the fact that the N is so high (and C so low) in the CNO-enhanced group indicates that there has been very strong hydrogen burning on top of this. This is supported by the low $\rm^{12}C/^{13}C$ and very low [C/N] (see Table~\ref{tab:table4}), both indicating equilibrium CN(O) cycling. To reach CNO equilibrium typically takes about $10^{4}$~years \citep{Caughlan1965}. This puts a time constraint on the minimum burn time for the event that gave rise to the chemical pattern in Group 3. Sodium is also enhanced, which is also expected for hot hydrogen burning. This suggests the burn temperature was high enough to activate the NeNa chain at around 50-60~MK \citep{Arnould1999}. We note that this chemical pattern is very similar to that of Hot Bottom Burning (HBB) on top of third dredge-up (TDU) in AGB stars (e.g., \citealt{Karakas2014}; see discussion in Section~\ref{sec:agb}). Any pollution scenario would need to reproduce HBB conditions, at least for Group~3 stars. This may be achieved, for example, by stellar mergers, as discussed below (Section~\ref{sec:OLRC_merger}). 

    \begin{table}[]
    \caption{Comparison of our classifications of the CDGs. Approximate averages and ranges are given. See text for definition of each sample/group.}
    \label{tab:table4}
    \begin{center}
   \hspace*{-2.1cm} \begin{tabular}{c|c|c||c|c}
    \hline
                                          & Group 1 & Group 2 & Group 3a & Group 3b \\ 
                                          & (normal-  & (over-  &    &    \\
                                          & luminosity-  & luminous-  &  (TESS)  &  (Lit.)  \\
                                           & RC)  & RC)  &    &    \\
                                         
    \hline
    \hline
    Lum. (L$_{\odot}$)                    & $50-80$ & $95-150$& $120-155$& $110-760$\\
    Mass (M$_{\odot}$)                    &$1.2-2.2$&$1.4-2.4$& $1.6-3.0$& ?        \\
    {[}Na/Fe{]}                           & $0.0$   & $+0.3$  & $+0.4$   & $+0.4$   \\
    {[}CNO/Fe{]}                          & $0.0$   & $+0.1$  & $+0.2$   & $+0.4$   \\
    Li-rich frac.                         & $90\%$  & $0\%$   & $85\%$   & $40\%$   \\
    {[}C/Fe{]}                            & $-0.5$  & $-0.7$  & $-1.3$   & $-1.1$   \\
    {[}N/Fe{]}                            & $+0.5$  & $+0.6$  & $+1.1$   & $+1.3$   \\
    {[}O/Fe{]}                            & $0.0$   & $+0.1$  & $+0.1$   & $+0.2$   \\
    {[}C/N{]}                             & $-1.1$  & $-1.3$  & $-2.4$   & $-2.4$   \\
    $\rm^{12}C/^{13}C$                    & ?       & ?       & $6$      & $5$      \\
    {[}Fe/H{]}                            & $-0.1$  & $-0.3$  & $-0.2$   & $-0.2$   \\
    \hline
    \end{tabular}%
    %}
    \end{center}
    \end{table}

    Turning to the Group~1 and 2 stars, they are distinguished by having [C+N+O/Fe] approximately unchanged from their initial scaled-solar values. This indicates that, unlike Group~3, the polluting material was not enriched in He-burning products through core dredge-up. The degree of N-enhancement/C-deficiency depends on the amount of burning that the material underwent (along with any dilution that may have occurred if the pollution from from an external source). Given their high [N/Fe] and low [C/Fe], these stars show signs of substantially progressed CN(O) burning. We do not have $\rm^{12}C/^{13}C$ for them, but the [C/N] is low (Table~\ref{tab:table4}). However, as shown by \cite{Clayton1983} we expect [C/N]~$\approx -2.5$~dex in equilibrium CN(O) burning at around $30-50$~MK, which is much lower than the observed $-1.1$ to $-1.3$~dex in these stars. Thus it appears that equilibrium was not quite reached during the burning that gave rise to the patterns in these stars. Since it takes about $10^{4}$~years to reach equilibrium \citep{Caughlan1965}, this puts a time constraint on the maximum burn time for the Group 1+2 chemical patterns. The [C/N] ratio is higher in Group 1, which could be due to a lower-temperature burn. Further, Na is not enhanced in Group 1 while it is in Group 2. This suggests that the burn temperatures were around $20-40$~MK (\citealt{Adamczak2013}) in Group 1 but around $50-60$~MK in Group 2. This may have allowed Li to survive in the Group 1 stars, as opposed to Group 2 (Table~\ref{tab:table4}; also see Li discussion in Section~\ref{sec:link_to_lithium}).

\subsection{Revisiting previous theories on the origin of C-deficient stars} \label{sec:revisiting}

    Before the current study, the general picture of the nature of wGb/CDGs was that:
    \begin{itemize}
        \item Most have uncertain evolutionary status (SGB, RGB, RC, EAGB).
        \item They are intermediate-mass stars ($M \approx 2.5~\rm M_{\odot} - 5.0~\rm M_{\odot}$).
        \item They may mainly be thick-disk objects given their large distances from the Galactic plane.
        \item All have N over-abundances.
        \item Some are enhanced in Li.
        \item Most are enhanced in Na.
        \item All those measured have low $\rm^{12}C/^{13}C$ ratios.
    \end{itemize}
    
    For seven decades the precise evolutionary state of the CDGs/wGb stars has been uncertain. This is due to the fact that they are found in a very crowded area of the HRD where it is difficult to distinguish between various evolutionary phases (SGB, RGB, RC, EAGB). 
    
    By taking advantage of the Kepler asteroseismic data our study shows conclusively that almost all (14/15) of our sample of stars are in the RC (core He-burning) phase. For our previously-known CDG sample with  data, for which we have only $\nu_{max}$, we find they are also consistent with being RC stars (Figure~\ref{fig:fig10}). Considering the wider literature sample of CDGs, we find that many of them are consistent with RC luminosities but there is also a subset that has much higher luminosities (black dots in Figure~\ref{fig:fig10}).
    
    The fact that there are very few stars below the RC luminosity -- we have one single example out of 44 stars (15 Kepler CDGs + 29 known-CDGs; Figures~\ref{fig:fig1} and \ref{fig:fig10}) -- means that the CDG chemical signature could not have arisen before the RGB bump (since the bump has roughly the same luminosity at the RC). Further, if almost all CDGs are RC stars (we require more seismic data to determine this), the pollution event could not have occurred before the RGB tip (except in a merger scenario, see Section~\ref{sec:OLRC_merger}). This important new finding helps constrain theories of the origin of the CDGs. Theories that suggest that the chemical pattern was imprinted during the PMS or MS are not supported by this new evidence. This applies to self-enrichment theories (e.g. self-enrichment through rapid rotation; \citealt{Adamczak2013}) and also external pollution theories, for example where the chemical pattern is thought to reflect the birth composition of the stars, whereby they could have been born polluted or have been polluted in their infancy by more massive stars that would have ejected CNO cycle processed material \citep{Palacios2016}. Previous studies did note that these theories did have their weaknesses, for example, there were no known MS stars with the same chemical pattern as the CDGs that could have been progenitors. We note that we also did not find any MS or SGB C-deficient stars (Figure~\ref{fig:fig1}), further suggesting that the pollution event doesn't happen early in the evolution.
    
    Another possible theory in the literature is that CDGs are products of mass transfer or merger events \citep{Bond2019}. This is based on three pieces of evidence, (i) the systematically high distances from the Galactic plane as compared to normal red giants lying in the same location in the CMD, (ii) their apparent higher masses ($2.0 - 4.5~\rm{M}_{\odot}$), which might imply they are binaries or binary products, and (iii) a subset having high rotation rates. In contrast, for our sample (for which we have high-quality seismology) we find that (i) there is no preference for the Galactic location of these CDGs (Figure~\ref{fig:fig6}; Table~\ref{tab:table1}), and (ii) our CDGs are predominantly low-mass stars ($1.2 - 2.3~\rm{M}_{\odot}$; Figure~\ref{fig:fig5}). Due to these findings, it can be seen that the \cite{Bond2019} theory of CDGs belonging to a special stellar population may not apply to our sample. Although the masses of our sample are relatively low, we note that they still appear biased towards higher masses, just not as strongly. We consider the merger scenario for our sample in light of our new constraints in Sections~\ref{sec:OLRC_merger} and~\ref{sec:NLRC_merger}.
    
    As a caveat we note that in Section~\ref{sec:luminosity_bimodality_section} we reported seismic masses for 6 stars from the literature. Five out of 6 of these stars were in the \cite{Bond2019} study. As mentioned in Section~\ref{sec:luminosity_bimodality_section}, there is a tension between our seismic masses, which are generally low, and our non-seismic masses which are generally higher (Figure~\ref{fig:fig9}). This adds some uncertainty around the formation scenario(s), for this subset of stars.

\subsection{Possible explanations for the origin of Group 2 and Group 3 CDG stars based on our new findings}

\subsubsection{Over-luminous-RC CDGs (Group 2): Helium subflashing stars?} \label{sec:olrc_subflash}

    At the tip of the RGB, low-mass stars go through the core helium flash. After this they undergo a series of weaker He-flashes as they descend down to the RC luminosity (\citealt{Bildsten2012}; \citealt{Raghu2021}), where they then spend $\simeq 80-100$~Myr. As mentioned above, two out of the 11 stars (18\%) for which we have $\Delta\Pi_{1}$ measurements are in the helium subflashing phase according to \cite{Mosser2014}. This is 4.5 times the expected percentage (3.8\%) found in our background sample of RC stars. Also, both stars are over-luminous-RC CDGs (Table~\ref{tab:table4} and Figure~\ref{fig:fig10}), and in the seismic diagram of Figure~\ref{fig:fig11} almost all of the over-luminous-RC CDGs are found in the same region as the subflashing stars. Thus it is tempting to associate our over-luminous-RC CDGs with stars currently undergoing helium-burning subflashes.

    The time spent in subflashes (with convective He-burning shells) is very short ($\simeq 10^{4}$~years; \citealt{Bildsten2012,Raghu2021}), so it is very unlikely to find stars in this phase. If these stars are in the subflashing phase then they will soon evolve to the RC. Since our over-luminous-RC CDG stars are C-deficient, this would mean that we would expect to see large numbers of C-deficient RC stars, which we do not. Given this, it appears that our over-luminous CDGs cannot be subflashing stars. Their location in the seismic diagram (Figure~\ref{fig:fig2}) is likely degenerate with other phases of evolution -- or different stellar structures, as discussed in the next subsection.

    Another argument against these stars being subflashing stars is that, theoretically, it is expected that any mixing at the core helium flash would result in an \textit{increase} in carbon at the surface (e.g., \citealt{Deupree1987,Izzard2007,Mocak2009}). However it has been suggested that Li-rich stars may provide evidence for flash-induced mixing without CNO enrichment (Section~\ref{sec:link_to_lithium}).

\subsubsection{Over-luminous-RC CDGs (Group 2): Merger products?}  \label{sec:OLRC_merger}

    Our results show that our over-luminous-RC CDGs (Group 2 in Table~\ref{tab:table4}) stand out in a number of ways:

    \begin{itemize}
        \item Critically, they are more luminous than expected for low-mass RC stars, by a factor of about 2 ($\approx 60~\rm{L}_\odot$ vs $120~\rm{L}_\odot$; Figure~\ref{fig:fig11}).
        \item They have higher average masses than our background RC sample. In particular, there are no stars below $1.35~\rm M_{\odot}$ (Figure~\ref{fig:fig11}).
        \item The majority of these stars also have abnormally large seismically-measured radii (middle panel of Figure~\ref{fig:fig11}) with the 4 largest stars (out of 6) having an average radius of $17.0 \pm \rm{0.8}~R_\odot$ which is $\simeq 5.5~\rm{R}_{\odot}$ larger than the background sample of RC stars ($R_{avg} =\rm 11.5~R_\odot$ in the mass range of interest). 
        \item These stars have more extreme chemical patterns than the normal-luminosity-RC CDGs (Group 1 in Table~\ref{tab:table4}). In particular, they show Na enrichment ([Na/Fe]$_{\rm avg} = +0.3$~dex) whereas the normal CDGs have scaled-solar Na. For C and N our overluminous-RC-CDGs are 0.2~dex more C-deficient  and 0.1~dex more N-rich  (Table~\ref{tab:table4} and Figure~\ref{fig:fig10}).    
        \item They also stand out in the seismic diagram of Figure~\ref{fig:fig11}, well away from the normal RC distribution. 
    \end{itemize}

    The combination of their relatively high masses (compared to normal RC stars), higher luminosities and radii (for their masses) suggests that these stars may be merger products. This was proposed by \cite{Bond2019} for their sample of CDGs. Our seismic information strengthens the fact that these stars are quite different to normal RC stars. In Figure~\ref{fig:fig11} we have 5 of the \cite{Bond2019} stars (with  seismic parameters). They appear to be equivalent to our over-luminous-RC CDGs, although their C depletion is even stronger (and N, Na more enhanced). The \cite{Bond2019} sample being more chemically extreme may be due to selection bias through how they were discovered. They were classified based on the weak CH-band in the spectra recorded in objective prism plates (\citealt{Bidelman1951}; \citealt{Bidelman1973}).

    As discussed in Section~\ref{sec:chemical_signatures}, the chemical pattern seen in these stars reflects previous hydrogen burning through the CN (or CNO) cycle, and the presence of Na indicates that the NeNa chain was also running. This suggests the burn temperature was around 50-60~MK \citep{Arnould1999}. \cite{Adamczak2013} also noted that some CDGs have unusually high Li abundances indicating Li production presumably through the Cameron–Fowler mechanism. On the other hand, our sample of overluminous-RC CDGs (Group 2 in Table~\ref{tab:table4}) does not contain any very Li-rich stars (all stars for which we have spectra have A(Li) < 1.8~dex; Table~\ref{tab:table3}). This is in stark contrast to our normal-luminosity-RC CDG sample (Group 1) of which $\simeq 86\%$ are very Li-rich (6/7 stars; we have spectra for 7 out of 8 of this sample; Table~\ref{tab:table3}).

    If we accept the merger scenario for these overluminous-RC CDGs and combine it with our result that virtually all of the CDGs are at the RC luminosity or brighter, the burning must have happened during or after the merger event. That is, since the chemical pattern is not seen in earlier phases of evolution, the most simple explanation is that it arises as a result of the merger event.

    Interestingly, in the low-mass merger models of \cite{Zhang2013}, there is a case in which CN(O) and NeNa nucleosynthesis does indeed occur (their model B2). Although this model did not match the chemical patterns of the stars they were studying (early-type R and J stars; carbon-rich), they suggested that this type of merger could help explain the globular cluster abundance anomalies, which show a similar chemical pattern to the CDGs. Their model produces stars with surface abundances [C/Fe]$ = -1.15$~dex, [N/Fe]$ = 0.88$~dex, [O/Fe]$ = -0.10$~dex, and [Na/Fe]$ = 1.42$~dex. Qualitatively this appears to match the abundances of the over-luminous CDGs. They mention that Li is at first produced, but quickly destroyed. Their model B2 is a merger between a helium white dwarf ($0.2~\rm M_{\odot}$) and an RGB star with a He core mass of $0.3~\rm M_{\odot}$. This is a common formation channel as shown by binary population synthesis (e.g., \citealt{Izzard2007}). In the B2 model case of \cite{Zhang2013} accretion is from the HeWD onto the RGB helium core.

    \begin{figure}[htp]
    \centering
    \includegraphics[width=1.\linewidth]{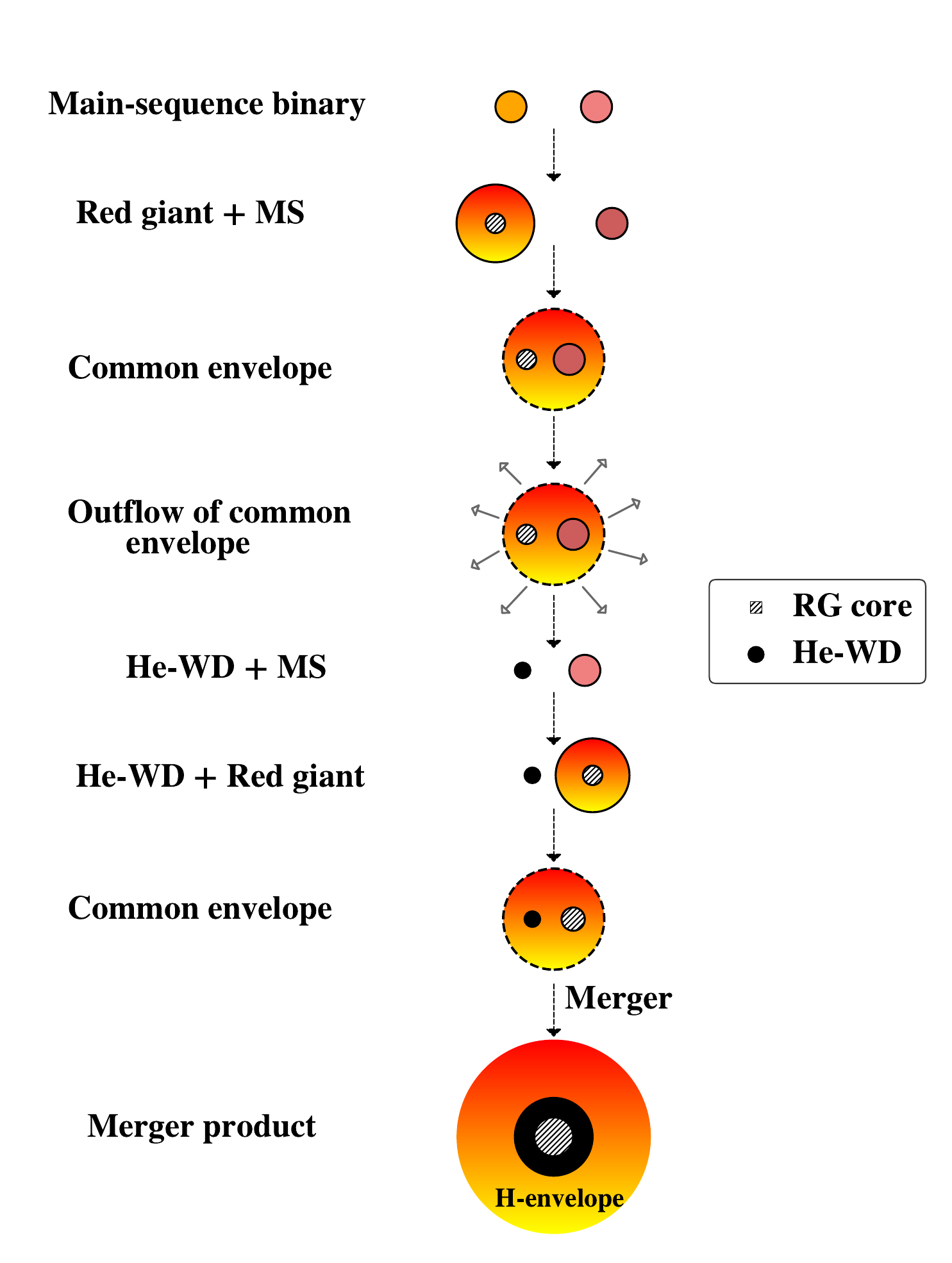}
    \caption{Schematic representation of the possible steps in a HeWD plus RGB star merger leading to the formation of a red clump star with a carbon-poor and nitrogen-rich surface. This illustration is based on Figure~5 in \cite{Postnov2014} and Figure~2 in \cite{Zhang2013}.}
    \label{fig:fig14}
    \end{figure}

    The full evolutionary picture is shown in Figure~\ref{fig:fig14}. Once a HeWD + RGB binary has formed, the RGB star eventually expands and overflows its Roche lobe, leading to a common envelope event. It is assumed that the HeWD merges with the RG core, which means that spiral-in must occur faster than envelope ejection. The resulting merged star has a He-rich core of $0.5~\rm M_{\odot}$ with a H-rich envelope. The \cite{Zhang2013} B2 model accretes enough envelope mass to make a total merged-star mass of $2~\rm M_{\odot}$. Evolving the model forward, it underwent some strong burning at the base of the convective envelope, which reaches right down to the H-burning shell (their Figure~8). The study did not report the temperature of the envelope burning however it can be expected to be in the range 50-70~MK judging from the nucleosynthesis that occurs. The burning lasts about 3 million years, at which point it undergoes a series of He-shell flashes before settling into convective-core helium burning (RC), where it remains for $\simeq 50$~million years. It is this RC phase that we associate with our observed sample of over-luminous-RC CDG stars -- they have higher than normal luminosities, a similar chemical pattern, and are core-helium burners (RC, as identified through Kepler asteroseismology, see Section~\ref{sec:seismic_diagram}). Although there is not an exact quantitative match between our abundances and this particular merger model, it can be expected that a change in model parameters (e.g., HeWD mass, RG core mass), will lead to a variation in burning products due to different thermal structures.

    As for the previously-known wGb stars for which we have some  seismology (Group 3a in Table~\ref{tab:table4}; open squares in Figures~\ref{fig:fig10} and \ref{fig:fig11}), they appear to be members of our over-luminous-RC group, with slightly more extreme chemical patterns (Section~\ref{sec:luminosity_bimodality_section}), so they also fit this merger scenario.

    In Figure~\ref{fig:fig10} we also show an extended literature sample of CDGs (Group 3b in Table~\ref{tab:table4}), for which we do not have asteroseismic data. These stars show similar extreme abundance patterns as the known sample for which we have  data. However they show a wide distribution in luminosity, reaching far above the RC luminosity (and above the overluminous-RC). One interpretation of these much brighter stars is that they are evolved versions of the RC stars (early-AGB). However, if this were the case then we would expect to see bright CDGs with C-depletion similar to the normal-luminosity and overluminous-RC CDGs (i.e. lying directly above these stars in the left-most plot of the lower panel of Figure~\ref{fig:fig10}), but we do not observe this. Lacking data to confidently assign evolutionary status, we speculate that the bright extension of CDGs are also merger-product RC stars, and the increase in luminosity is due to increasing mass. More data is required to test this hypothesis. We note that these stars may be over-represented in the literature sample due to their brightness.

    We conclude that a merger between a helium white dwarf and a red giant can explain the overluminous-RC CDGs (Group 2), and possibly the bright extension seen in the literature sample of wGb stars (Group 3).

\subsubsection{Previously-known-wGb stars (Group 3): AGB mass transfer pollution?}
\label{sec:agb}

    As mentioned in Section~\ref{sec:characterisation_of_cdgs}, AGB stars that are undergoing HBB on top of TDU can qualitatively reproduce the chemical pattern in the Group 3 stars. In particular, stars of $\simeq 6~\rm{M}_{\odot}$ produce Na and Li (the Li-rich phase is transient;  \citealt{Karakas2014}), as well as increasing [C+N+O/Fe], as observed in these CDG stars (Table~\ref{tab:table4}). However, the luminosity range of our sample ($\sim 10^{2} - 10^{3}$~L$_{\odot}$) is not compatible with HBB AGB stars ($\sim 10^{4}$~L$_{\odot}$). Also, the CDG masses are not as high as $6~\rm{M}_{\odot}$. Therefore, if the material did originate in an AGB star, it must have been added to the surface of these stars through binary mass transfer.

    A mass-transfer scenario would imply the current Group 3 CDGs should be in binary systems. We checked for binarity in the literature, consulting multiple catalogues (\citealt{Mason2001,Dommanget2002,Kovaleva2015,Kareem2021}) and found 12 giants either in multiple systems or in a binary system. This is 41$\%$ of the 29 known-CDGs, which is broadly consistent with the expected binary fraction in this mass range ($\sim 48\%$; \citealt{Parker2014}). Thus there appears to be no bias towards a high binary fraction that would expect from this scenario. We note however that this result may be affected by incomplete data. A dedicated binary survey of these stars would be needed to be certain of the true binary fraction. If the AGB binary mass-transfer scenario were correct, we would also expect the primary (donor) stars to have evolved to the WD by now, so we would expect to see Group 3 stars with WD companions. We only find three of the 29 Group 3 stars with evolutionary phase identifications in the literature. All of these are identified as MS-MS binaries \citep{Kareem2021}, at odds with the scenario. More data on binarity is required to properly test the AGB binary mass-transfer theory.

\subsection{Possible explanations for the origin of Group 1 CDG stars based on our new findings}

\subsubsection{Normal-luminosity-RC CDG stars (Group 1): Core-flash mixing?} \label{sec:NLRC_coreflashmixing}

    Our sample of normal-luminosity-RC CDG stars (Group 1 in Table~\ref{tab:table4}) have luminosities consistent with the background sample of RC stars (hence the name; Figure~\ref{fig:fig11}). They also mostly have normal RC radii although there is one outlier that has a large radius. They have less extreme chemical patterns, being less C-deficient and less N-enhanced. A striking difference with the overluminous sample is that they do not have any Na overabundance -- they all have scaled-solar Na. This suggests the burning that led to the chemical pattern of these stars was not at such high temperature, with just the CN cycle operating ($\simeq 20-40$~MK; \citealt{Adamczak2013}). The masses of the normal-luminosity-RC CDGs are similar to the overluminous sample but biased to slightly lower masses (Figure~\ref{fig:fig10}; Table~\ref{tab:table4}). Despite this, the sample is also missing the lowest mass stars as compared to the background RC sample, with no stars of mass less than $1.2~\rm{M}_{\odot}$. This could be evidence of a merger scenario, discussed in Section~\ref{sec:NLRC_merger}. With respect to seismology, they primarily follow the normal RC distribution (see seismic diagram in Figure~\ref{fig:fig11}). Interestingly, most of this group is very Li-rich. Six out of the 7 stars ($\simeq 86\%$) for which we have spectra are Li-rich (Table~\ref{tab:table3}). We discuss the remarkable number of Li-rich CDGs in Section~\ref{sec:link_to_lithium}.

    The fact that CDGs are generally not found at luminosities below the RC is suggestive of a pollution event near the RGB tip. Given the mass range of our Group 1 stars (Table~\ref{tab:table4}), they should all have gone through degenerate ignition of He in the core -- the core He flash (CHeF). Although envelope pollution has been suggested at the CHeF (e.g., \citealt{Deupree1987,Izzard2007,Mocak2009}), it is expected that any mixing at the core helium flash would result in an \textit{increase} in carbon at the surface (e.g., \citealt{Deupree1987,Izzard2007,Mocak2009}), incompatible with the C-deficient stars. However a `weaker' pollution event has been suggested at the CHeF to explain Li-rich giants (\citealt{Kumar2011}), which are primarily RC stars too, and also to explain `Li-enhanced' RC stars (\citealt{Kumar2020}; \citealt{Schwab2020}). One might imagine a flash-induced mixing scenario between these two extremes, whereby CN(O) burning operates, producing N-enhancement and C-deficiency. As mentioned, the fact that there is no Na enhancement in these stars means that the hypothesized extra-mixing would not have to reach very high burning temperatures. As far as we are aware CHeF extra-mixing in this regime has not yet been investigated. Finally, since there are so few CDGs ($0.15\%$ of our sample; Section~\ref{sec:sample_selection}), this proposed extra-mixing event would only occur in a very small fraction of stars.

\subsubsection{Normal-luminosity-RC CDG stars (Group 1): Mergers?} \label{sec:NLRC_merger}

    As mentioned above, the normal-luminosity RC CDGs are low mass but show a bias against the lowest masses relative to the background RC sample (Figure~\ref{fig:fig11}). This is similar to our overluminous sample (Group 2 in Table~\ref{tab:table4}), for which we concluded the merger scenario to be the most likely. In particular, for the normal-luminosity sample, there are no stars below $\simeq 1.2~\rm M_{\odot}$, and the average mass of the stars is $1.5~\rm M_{\odot}$. This could be the signature of a merger product population since mergers tend to produce more massive stars.

    Moreover, in the seismic diagram (Figure~\ref{fig:fig11}c), although on the whole, this sample overlaps the background RC sample, it does show significant dispersion. Half of the sample (4/8 stars) are on the extremes of the background distribution. Further, two of them  clearly fall above the general trend (having relatively high ${\rm \nu_{max}}^{0.75}/ \Delta \nu$ and $\nu_{max}$), near the position of the overluminous-RC CDGs. This is further suggestive of the merger scenario, and the dispersion may be related to variation in the progenitors. We note that these stars do not appear to be stripped stars as identified by \cite{Li2022} (orange symbols in Figure~\ref{fig:fig11}).

    We suggest that the fact that these stars are not overluminous does not necessarily rule out the merger scenario. This is because there may be a bifurcation in core-mass post merger; lower-mass mergers will undergo the CHeF, whereas higher-mass mergers may not (depending on the resultant merged core mass). Alternatively, other merger scenarios or physics (e.g., \citealt{Zhang2013}) could produce the variation.

    As mentioned, the chemical signature of these normal-luminosity RC CDGs is relatively limited, compared to the overluminous sample. This could be consistent with the lower masses of the normal-luminosity stars (on average $0.3~\rm{M}_{\odot}$ lower), such that lower-mass mergers are likely to result in lower-temperature burning which would give less extreme chemical patterns.

\subsection{Link to the Li-rich giants} \label{sec:link_to_lithium}

    Considering the rarity of Li-rich giants ($\simeq 3\%$ on the RC; \citealt{Kumar2020}), there are a remarkable number of Li-rich C-deficient stars. In our sample for which we have spectra, $50\%$ (6/12 stars) have A(Li)~$> 1.8$~dex (1.8~dex is our detection limit; Section~\ref{sec:lithium_result_section}). In our overluminous literature sample with spectroscopic data (black open squares in Figure~\ref{fig:fig11}) about  $80\%$ (5/6 stars) are Li-rich (also see Table~\ref{tab:table4}).

    We compiled a comprehensive literature sample of previously-known CDGs for which Li abundances are available (\citealt{Adamczak2013}; \citealt{Palacios2016}) and find that $\sim 50\%$ of those stars (14/29 stars) are Li-rich (A(Li)~$> 1.5$~dex). Clearly there is an extremely strong bias towards C-deficient giants also being Li-rich giants. This suggests that they may be related in some way, so formation scenarios for the CDGs may also apply to the Li-rich giant population (and vice-versa).

    As discussed above, we have two main theories for our CDGs: CHeF mixing and low-mass stellar mergers. For our overluminous-RC CGDs we believe the merger scenario is the most likely, and for the normal-luminosity-RC CDGs we cannot differentiate between the CHeF-mixing and merger scenarios.

    Li-rich giants are generally (normal-luminosity) RC stars (\citealt{Kumar2011}; Figure~9 of \citealt{Zhang2020}; \citealt{Raghu2021}). That both Li-rich giants and CDGs are primarily RC stars is another striking parallel between these two chemically-peculiar populations, only just revealed by the current study.

    As we suggested in Section~\ref{sec:NLRC_coreflashmixing} for our CDGs, it has also been suggested that CHeF mixing may be responsible for the Li enrichment in Li-rich giants (\citealt{Kumar2011,Kumar2020, Schwab2020}). 
    
    The merger scenario has also been suggested for Li-rich giants by \cite{Zhang2020} who provide population synthesis models. \cite{Yan2021} further investigate this merger scenario. They are able to explain most features of the Li-rich giants, however, they cannot explain the nitrogen enhancements in many of the stars (see Figure~2 in their Extended Data section). Nitrogen enhancement implies that the material has experienced CN(O) burning, so we would expect C to have been depleted in some of the Li-rich giants, particularly the most N-rich stars. To check this, in Figure~\ref{fig:fig13} we compare the C and N abundances (and masses) for a sample of Li-rich giants to various samples of C-deficient giants. Indeed, taking Li-rich giants with strong N-enhancement ([N/Fe]~$ > +0.4$~dex), we find $\sim 85\%$ of them to be C-deficient giants as per our definition given in Section~\ref{sec:definition} ([C/Fe]~$< -0.4$~dex). In fact, 6 of our CDGs are included in the Li-rich study of \cite{Yan2021}. This again shows the significant overlap in Li-rich and C-deficient giants, which can be seen visually in Figure~\ref{fig:fig13}.

    Li-rich giants have low $\rm^{12}C/^{13}C$ ratios, being mostly in the range 5-10 (although in some cases it is close to 30; \citealt{Kumar2011}). This compares well with the CDGs for which $\rm^{12}C/^{13}C$ has been measured, which also show low values (Table~\ref{tab:table4}). As mentioned above, we do not have $\rm^{12}C/^{13}C$ for our Group 1 and Group 2 stars but their very low [C/N] indicates they have also undergone significant burning, albeit not to equilibrium (Section~\ref{sec:chemical_signatures}). Thus it is expected that their $\rm^{12}C/^{13}C$ must be fairly low\footnote{Also see the C-poor Li-rich giant in \citealt{Silva2014} which has an upper limit for $\rm^{12}C/^{13}C$ of 20.}.

    Given the aforementioned similarities/overlap between Li-rich giants and the CDGs, we speculate that we are seeing a spectrum of merger products, varying in progenitor mass.
    In Table~\ref{tab:table4} we see that Li-rich giants appear in Groups 1 and 3, which we have identified as having distinct chemical enrichments in CNO and Na. This shows that Li is not the best tracer for identifying chemically peculiar stars. This is likely due to its easily-produced and easily-destroyed nature. We note that the majority of Li-rich giants do not show signs of CNO burning in their envelopes (Figure~\ref{fig:fig13}). If we extend the merger hypothesis to these stars, it may indicate lower-temperature burning and therefore lower-mass progenitors. Alternatively, their Li may have been formed at the core helium flash \citep{Kumar2020,Schwab2020}.

\subsection{Globular cluster link}

    As noted by \cite{Zhang2013}, the chemical pattern of their B2 merger model is similar to the abundance anomaly patterns found in second-generation globular cluster populations: Low C, high N, and enhanced Na. We also identified the B2 model as a possible match for many of our CDGs, since this pattern also qualitatively matches them. However, as with the \cite{Zhang2013} model, the CDGs do not match the oxygen depletion seen in GCs \citep{Gratton2000}, which gives rise to the Na-O anti-correlation. As speculated by \cite{Zhang2013}, this may be a metallicity effect -- metal-poor CDGs (or merger models) may deplete O. This is an enticing scenario for GCs since it is expected that stellar interactions are relatively common in such dense stellar environments, with mergers being a key binary destruction pathway (see e.g., Figure~4 in \citealt{Ivanova2005}).

\section{Conclusion}\label{sec:conclusion}

    In this study, we addressed the long-standing questions related to the mass and evolutionary phase of the carbon-deficient giants (CDGs) with the help of asteroseismology (Kepler and TESS~missions), spectroscopy (APOGEE and LAMOST surveys), and astrometry (Gaia). We list our wide-ranging results in Section~\ref{sec:results_overview} (also see Table~\ref{tab:table4}). We briefly summarise the most important points here.

    We found 15 newly-identified CDGs in the Kepler field. As a fraction of our Kepler-APOGEE sample, CDGs represent only $0.15\%$ of the stars and hence are very rare.

    For the first time, we unambiguously identify the evolutionary state of CDGs. Remarkably, we found 93\% of our sample to be in the red clump (RC) core He-burning phase. This puts strong constraints on formation scenarios. The lack of MS, SGB and RGB carbon-deficient stars rules out theories that posit the chemical pollution occurred early in CDG evolution, be it through self-pollution or external pollution scenarios.

    In contrast to previous literature, where wGb stars were considered to be of intermediate mass ($M \approx 2.5~\rm M_{\odot} - 5.0~\rm M_{\odot}$) based on their position on the HRD, the Kepler-based asteroseismic masses reveal that our sample of CDGs are primarily low-mass stars ($M \lesssim$ 2~M$_{\odot}$). We also determined asteroseismic masses for a small sample of previously-known wGb stars, finding that their seismic masses are also primarily low. This finding puts further constraints on any formation scenario.

    We find definite demarcations in the chemical patterns of CDGs, which enabled us to split them into three groups, in increasing order of degree of pollution. The first, Group 1, is characterized by  non-equilibrium CN-cycle burning pattern, whilst Group 2 is similar but slightly more extreme, but with an additional enhancement of Na. The third Group has an even more extreme chemical pattern and is distinct in that its pattern shows dredge-up of He-burning products has occurred, with strong hydrogen burning on top of this. The chemical patterns through the three groups also suggest an increasing temperature of hydrogen burning.

    We find a bimodality (roughly 60:40) in luminosity within our sample, with one group (Group 1) having normal RC luminosity and the other group (Group 2) being about a factor of two more luminous than expected for their masses ($\approx 60~\rm{L}_\odot$ vs $120~\rm{L}_\odot$). We concluded that stars in the more luminous group are likely merger products, having formed through the merger of a HeWD and an RGB star. One of the low-mass merger models of \cite{Zhang2013} is a good qualitative match for these stars. Previously-known wGb stars for which we have TESS seismology are also over-luminous, and we suggest they were formed through mergers as well. They do have more extreme chemical patterns and, importantly, are enriched in CNO elements, most likely from the dredge-up of He-burning products. In the merger scenario, this suggests that the progenitor stars would have been more massive, leading to more extreme mixing and burning \citep{Zhang2020}.

    For the normal-luminosity-RC CDGs (Group 1), we cannot distinguish between the two remaining formation scenarios. The first is a possible pollution event near the end of the RGB, likely at the core He-flash. This has also been suggested for the Li-rich giants (\citealt{Kumar2011,Kumar2020,Schwab2020,Raghu2021}). Interestingly, about 90\% of our normal-luminosity-RC CDGs are also Li-rich giants. The second suggested scenario is that these stars are  low-mass merger products like the over-luminous groups, just with less extreme chemical pollution, which might indicate less massive progenitor masses.

    A significant fraction of our CDGs (50\%) are also Li-rich giants. This is 17 times higher than the expected fraction of 3\%. Thus there is a strong overlap between CDGs and Li-rich giants. That said, there is variation between our three proposed CDG groups -- one group is $90\%$ Li-rich, the second has zero Li-rich members, and the third group is $50\%$ Li-rich ($85\%$ in our small TESS sample, which is biased to lower luminosities). As is well recognized, lithium is a special element -- it is easily produced but also easily destroyed. Thus it is not surprising to see this variation. In the merger models of \cite{Zhang2013} and \cite{Zhang2020} they also find Li to sometimes be produced and sometimes to be produced then destroyed.

    To summarise, based on our new data on carbon-deficient giants, we suggest the following scenarios for their formation, for the three groups we have defined:

    \begin{itemize}
        \item Group 1 (normal-luminosity-RC CDGs): Core He-flash pollution event, or low-mass mergers between HeWDs and RGB stars.
        \item Group 2 (over-luminous-RC CDGs): Higher-mass mergers between HeWDs and RGB stars.
        \item Group 3 (previously-known wGb stars, many over-luminous): Mergers between HeWDs and RGB stars, or binary mass-transfer from intermediate-mass AGB stars.
    \end{itemize}
    
    As can be seen, stellar mergers feature as a possibility across all groups. It is tempting to see the different groups of CDGs as being produced by mergers with different progenitor masses. More observations are required to confirm (or disprove) these proposed scenarios.

\section*{Acknowledgements}

 This study is supported by the National Natural Science Foundation of China under grant Nos. 11988101, 11890694, and 11873052 and National Key R\&D Program of China No.2019YFA0405500. S.M. acknowledges the support by CAS-TWAS Presidents fellowship for International Doctoral Students. S.M. thanks Raghubar Singh for helpful conversations. We thank Evgenii Neumerzhitckii for the use of his plotting routine. Funding for LAMOST (\url{www.lamost.org}) has been provided by the Chinese NDRC. LAMOST is operated and managed by the National Astronomical Observatories, CAS. This work has made use of data from the European Space Agency (ESA) mission \textit{Gaia} (\url{https://www.cosmos.esa.int/gaia}). We made use of the SIMBAD database and the VizieR catalog access tool, CDS, Strasbourg, France. S.W.C. acknowledges federal funding from the Australian Research Council through a Future Fellowship (FT160100046) and Discovery Projects (DP190102431 \& DP210101299). Parts of this research was supported by the Australian Research Council Centre of Excellence for All Sky Astrophysics in 3 Dimensions (ASTRO 3D), through project number CE170100013. This research was supported by use of the Nectar Research Cloud, a collaborative Australian research platform supported by the National Collaborative Research Infrastructure Strategy (NCRIS). S.W.C. thanks Carolyn Doherty for interesting and helpful discussions.

\end{document}